\documentclass[fleqn,10pt]{wlscirep}
\usepackage[utf8]{inputenc}
\usepackage[T1]{fontenc}
\usepackage{lineno}
\usepackage{arydshln}
\usepackage{multirow}
\usepackage{amsmath}

\title{LOMS.cz: A computational platform for high-throughput Classical and Combinatorial Judd-Ofelt analysis and rare-earth spectroscopy}

\author[1,2,*,+]{Jan Hrabovsky}
\author[3,4,+]{Petr Varak}
\author[5,6,+]{Robin Krystufek}
\affil[1]{Charles University, Faculty of Mathematics and Physics, Ke Karlovu 5, 121 16 Prague, Czech Republic}
\affil[2]{Department of Physics, New Mexico State University, MSC 3D, Las Cruces, 88003-8001, NM, United States}
\affil[3]{Department of Inorganic Chemistry, Faculty of Chemical Technology, University of Chemistry and Technology, Prague, Technicka 5, 166 28 Prague, Czech Republic}
\affil[4]{Institute of Photonics and Electronics of the Czech Academy of Sciences, Chaberska 1014/57, 182 00 Prague Czech Republic}
\affil[5]{Institute of Organic Chemistry and Biochemistry of the Czech Academy of Sciences, Flemingovo n. 2, Prague 6 16610, Czech Republic}
\affil[6]{Department of Physical and Macromolecular Chemistry, Faculty of Science, Charles University Hlavova 8, Prague 2 12843, Czech Republic}
\affil[*]{mail@janhrabovsky.cz}

\affil[+]{these authors contributed equally to this work}

\begin{abstract}

We present LOMS.cz (\textbf{L}uminescence, \textbf{O}ptical and \textbf{M}agneto-optical \textbf{S}oftware), an open-source computational platform that addresses the long-standing challenge of standardizing Judd-Ofelt (JO) calculations in rare-earth spectroscopy. Despite JO theory's six-decade history as the fundamental framework for understanding $4f\leftrightarrow4f$ transitions, the field lacks standardized computational methodologies for precise and reproducible parameter determination. LOMS integrates three key innovations: (1) automated computation of JO parameters, transition probabilities, branching ratios, and theoretical radiative lifetimes, (2) a dynamically expanding database of experimentally validated parameters enabling direct comparison between computed and empirical results, and (3) a novel Combinatorial JO (C-JO) analysis algorithm that systematically identifies optimal absorption band combinations to ensure reliable parameter extraction. As a proof-of-concept, we demonstrate how this computational framework enables rapid screening of spectroscopic parameters, allowing researchers to predict optical properties with enhanced reliability. By combining automated analysis with experimental validation through its integrated database, LOMS.cz establishes a standardized platform for accelerating the discovery and optimization of rare-earth-based photonic and optoelectronic materials.
  
\end{abstract}
\begin{document}

\flushbottom
\maketitle

\thispagestyle{empty}


The computational design and characterization of rare-earth (RE) materials represents a critical challenge in materials science, particularly given their essential role in modern technology. RE elements, especially their trivalent ions, exhibit unique electronic, magnetic, and spectroscopic properties that make them indispensable in various high-tech applications \cite{WYBOURNE2004_re_intro,Walsh2006,chen_2017_RE,liu_RE_spectroscopy}.  Within the industrial sector, RE ions are essential components in the manufacturing process of strong permanent magnets, which are used in electric cars, imaging devices such as the screen of smartphones/computers or as catalysts in chemical reactions\cite{Baolu_Zhou_Rare_2016,Gutfleisch_re_magnets,WYBOURNE2004_re_intro,sagawa_re_intro}. Furthermore, their luminescent properties are used for medical imaging as diagnostic tools, enhancing the capabilities of modern healthcare technologies\cite{Walsh2006,WYBOURNE2004_re_intro,dong_re_intro,thou_re_intro_animalimag}. The global market for RE-based products, reaching nearly $\$$2 trillion by 2012 (approximately 5$\%$ of global GDP), underscores their technological significance\cite{loms_market,loms_market2}.

\begin{figure}[h]
\centering
\includegraphics[width=11cm]{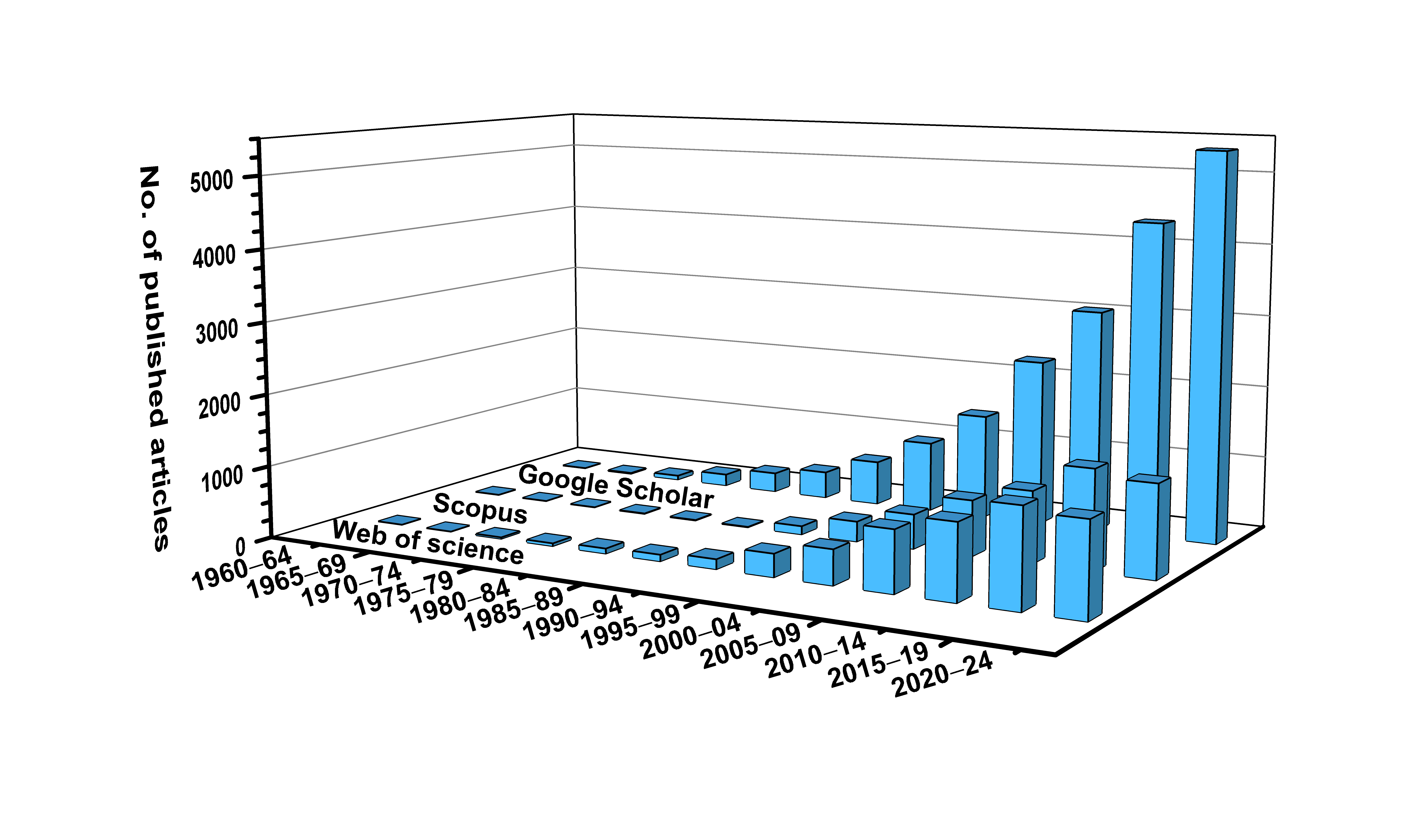}
\caption{\label{fig:loms_articles} Tracking of "Judd-Ofelt" expression within Google Scholar, Scopus and Web of sciences (WOS) scientific databases in 5-year intervals by July 2024.}
\end{figure}

A major challenge in RE materials research is the standardized analysis of their optical properties.  Despite extensive experimental knowledge of the spectroscopic properties of rare-earth ions, the correct mechanism of the intra $4f\leftrightarrow4f$ electronic transitions was only understood around the mid-20th century thanks to the advances in Racah's algebra and the enhanced computational capabilities brought by advancements in computer technology\cite{Walsh2006,HEHLEN2013_juddofelt,ciric2022_juddofelt}. Building on these previous accomplishments, B.R. Judd\cite{Judd1962} and G.S. Ofelt\cite{Ofelt962} independently introduced a theory in 1962 that describes the spectroscopic properties of rare-earth ions in various materials. These studies thus established the foundation for what later became known as the Judd-Ofelt (JO) theory, the first quantum-mechanical explanation of the electric-dipole induced $4f\leftrightarrow4f$ transition intensities in RE ions through the set of three JO parameters $\Omega_i (i=2,4,6)$. These parameters enable the prediction of  spectroscopic properties crucial for designing and optimizing photonic materials and devices, including transition probabilities $A(J',J)$, branching ratios $\beta (J',J)$, and theoretical luminescence radiative lifetimes, $\tau\textsuperscript{JO}\textsubscript{r}$.  
The exponential growth in JO theory applications, evidenced by over 19,000 publications indexed by Google Scholar by mid-2024 (see Fig.\ref{fig:loms_articles}), reflects three primary research directions: (1) theoretical advancement of JO parametrization methods\cite{Walsh2006,goldner_1996_jo,ciric2022_juddofelt,HEHLEN2013_juddofelt,smentek_2015_JO,sytsma_1989_jo}, (2) experimental characterization across diverse material systems\cite{loms_94}, and (3) integration of JO analysis into broader materials design strategies\cite{HEHLEN2013_juddofelt,ciric2022_juddofelt,CIRIC2019116749,CIRIC2019395_jo_intro}. It should be noted that the last two categories make up the majority of published works and primarily focus on the practical implementation of JO theory rather than its theoretical understanding. Over the years, various tools using the experiemntal emission spectra were developed to facilitate the practical application of JO theory, such as JOES \cite{Ciric2019} and LUMPAC \cite{Dutra2014} programs, enabling the computation of JO analysis for Eu\textsuperscript{3+} ion, or JOYSspectra \cite{Moura2021}, which allows the determination of JO parameters via theoretical calculations. However, a significant need remains for standardized experimental understanding and accessible computational tools to facilitate the practical implementation of JO theory and the reliable extraction of JO parameters from experimental data, especially using the optical absorption/transmission spectra. These include the complexity of selecting appropriate absorption bands for analysis, the challenge of ensuring reproducible parameter extraction, and the lack of systematic comparison between theoretical predictions and experimental results. To address these challenges, we developed Luminescence, Optics and Magneto-optics Software (LOMS), an open-source computational platform that automates and standardizes JO analysis. Our implementation of newly introduced Combinatorial JO (C-JO) analysis represents a significant advance in computational methodology, enabling systematic identification of optimal absorption band combinations for reliable parameter extraction. The LOMS.cz platform further introduces the first dynamic, systematically organized repository of JO parameters, facilitating direct comparison between computed and experimental results. Given a set of absorption spectra, the platform can fully automate the entire computational workflow, from initial parameter calculation through property prediction. As demonstrated below, LOMS.cz can systematically analyze spectroscopic parameters and predict optical properties while providing detailed uncertainty quantification through its comprehensive analysis approach. The platform enables researchers to efficiently evaluate materials properties and optimize RE-doped systems for specific applications, accelerating the development of next-generation photonic and optoelectronic materials through standardized computational analysis.

\section*{Results}
\subsection*{Method outline: Judd-Ofelt theory and Rare-earth ions}
To introduce JO theory and its implications, it is first necessary to define basic concepts related to the physics of rare-earth elements/ions, derivation of spectroscopic terms for RE\textsuperscript{3+} ground states as well as to know the position of other multiplets in energy diagram and other aspects required for in-depth spectroscopic description of solid. However, this section does not aim to provide an exhaustive mathematical treatment of JO theory and quantum mechanical descriptions, which are extensively detailed in original studies by Judd\cite{Judd1962} and Ofelt\cite{Ofelt962} or comprehensive works by Hehlen\cite{HEHLEN2013_juddofelt} and Walsh\cite{Walsh2006}. Instead, the primary goal is to present JO theory from an experimental perspective and introduce it to the broader scientific community. Presented outcomes are then introduced in the form of the interactive free-to-use computational online tool (www.LOMS.cz/jo) designed for calculating classical and combinatorial JO analysis and related parameters, facilitating accessibility and practical application of the theory. The calculated results can then be directly compared in the newly established JO parameter database on the same web platform (www.LOMS.cz/jo-database).

\subsection*{Rare-earth ions: spectroscopic properties and application}
The Rare-earth (RE) elements consist of seventeen chemical elements in the periodic table, including fifteen lanthanides (La, Ce, Pr, Nd, Pm, Sm, Eu, Gd, Tb, Dy, Ho, Er, Tm, Yb, Lu) along with Sc and Y. While rare-earth ions typically form trivalent cations, exceptions exist where divalent (Nd\textsuperscript{2+}, Sm\textsuperscript{2+}, Eu\textsuperscript{2+}, Dy\textsuperscript{2+}, Tm\textsuperscript{2+}, Yb\textsuperscript{2+}) and tetravalent (Ce\textsuperscript{4+}, Pr\textsuperscript{4+}, Tb\textsuperscript{4+}, Dy\textsuperscript{4+}) cations can also be formed. Rare-earth ions are widely used in electronics and in the production of magnets, catalysts, and photonics materials, with trivalent (RE\textsuperscript{3+}) cations being the most commonly utilized for these applications\cite{Baolu_Zhou_Rare_2016,Gutfleisch_re_magnets,WYBOURNE2004_re_intro,sagawa_re_intro,liu_RE_spectroscopy}. For this reason, the main focus will be on trivalent rare-earth cations with at least partially occupied 4f electron orbital and charge configuration of [Xe] 4$f$\textsuperscript{1$-$13}. Cations with fully filled (Lu\textsuperscript{3+}) or empty 4$f$-orbitals (La\textsuperscript{3+}, Y\textsuperscript{3+}) are not of spectroscopic interest as they do not allow any intra $4f\leftrightarrow4f$ transitions. However, despite the lack of inherent emission bands, Y\textsuperscript{3+}/Lu\textsuperscript{3+}/La\textsuperscript{3+} are substantial for various applications due to their capability of host matrix formation\cite{Hrabovsky_2021_yag_luag}. These ions thus provide a stable and inert surrounding for other activator ions from the RE ion group, such as Nd\textsuperscript{3+} (Nd:YAG lasers) or Ce\textsuperscript{3+} (Ce:YAG/LuAG-based light emitting diodes)\cite{Hrabovsky_2021_yag_luag,Baolu_Zhou_Rare_2016,Gutfleisch_re_magnets,WYBOURNE2004_re_intro,sagawa_re_intro,liu_RE_spectroscopy}. The comparable ionic radii and electronic structures allow them to form robust crystal lattices that can adopt a wide range of dopant ions to the order of tens of at.$\%$\cite{Hrabovsky_2024_RE_mapping}. This versatility makes them indispensable in the design of advanced phosphor materials (e.g. LED, solid state lasers), scintillators, and other luminescent materials for lighting, displays, and medical imaging technologies\cite{Baolu_Zhou_Rare_2016,Gutfleisch_re_magnets,WYBOURNE2004_re_intro,sagawa_re_intro,liu_RE_spectroscopy,Hrabovsky_2021_yag_luag,Hrabovsky_2024_RE_mapping}.

The primary benefit of optically active rare-earth ions with partially occupied 4$f$ electron orbitals is their spectroscopic stability within the host matrix regardless of whether the matrix consists of the above-described crystalline materials with or without the Y, Lu, La content, amorphous materials or special optical glasses. Emission bands from RE\textsuperscript{3+} ions in the host material closely match their intrinsic energies\cite{Walsh2006,liu_RE_spectroscopy,HEHLEN2013_juddofelt}, displaying narrow spectral lines across a broad wavelength range, from UV to MIR. In contrast, transition metals exhibit broader spectral lines due to the significant influence of the host matrix on their 3d shells\cite{Walsh2006}. This difference occurs because the 4f shells of lanthanides are partially shielded by their outer electron shells (5s and 5p) as is visible in Table \ref{tab:loms_reions}. This leads to a very weak interaction between these optical active electrons and the host matrix/surrounding ligand field. Perturbation of the local surrounding environment then affects the free RE\textsuperscript{3+} ion Hamiltonian ($H\textsubscript{F}$) and leads to the creation of Stark levels. The Hamiltonian of free RE\textsuperscript{3+} ion can be expressed using Eq.\ref{eq:jo_hamilt} as 

\begin{equation}
\label{eq:jo_hamilt}
H\textsubscript{F}=H\textsubscript{0}+H\textsubscript{C}+H\textsubscript{SO},
\end{equation}
 where the first term, H\textsubscript{0}, represents the nucleus-electron interaction and the kinetic energies of all the electrons, the second term is the coulombic repulsion between electrons, $H\textsubscript{C}$, and the last term describes the spin-orbit interaction, $H\textsubscript{SO}$, and thus coupling between the spin angular momentum and the orbital angular momentum. Previously mentioned interaction with the surrounding crystal/ligand field could then be expressed by adding another term representing the perturbation Hamiltonian, $V\textsubscript{LF}$, and form the perturbated free ion Hamiltonian for an ion in the host matrix as follows $H=H\textsubscript{F}+V\textsubscript{LF}$. For a more detailed description, please follow Refs.\cite{Judd1962,Ofelt962,Walsh2006,HEHLEN2013_juddofelt}.

\begin{table}[h]
\caption{Charge configuration of RE\textsuperscript{3+} ions, atomic number ($Z$), number of electrons in 4$f$ orbital ($n$\textsubscript{e}), total spin ($S$) and orbital ($L$) angular momentum , total angular momentum ($J$) and derived \textsuperscript{2\textit{S}+1}L\textsubscript{\textit{J}} ground spectroscopic term. }
\label{tab:loms_reions}
\fontsize{8pt}{8pt}\selectfont
\begin{tabular}{clclccclc}
\hline
Z & Element      & Symbol & RE\textsuperscript{3+} config.&$n$\textsubscript{e}& $S$& $L$&$J$& Ground\\
 & & & & & &  &&term\\
 & & & & & &  &&\\ \hline
58            & Cerium       & Ce     & {[}Kr{]}4$f$\textsuperscript{1}5$s$\textsuperscript{2}5$p$\textsuperscript{6}  &1  & 0.5 & 3                       &2,5 & \textsuperscript{2}F\textsubscript{5/2}            \\
59            & Praseodymium & Pr     & {[}Kr{]}4$f$\textsuperscript{2}5$s$\textsuperscript{2}5$p$\textsuperscript{6}  &2  & 1   & 5                       &4   & \textsuperscript{3}H\textsubscript{4}              \\
60            & Neodymium    & Nd     & {[}Kr{]}4$f$\textsuperscript{3}5$s$\textsuperscript{2}5$p$\textsuperscript{6}  &3  & 1.5 & 6                       &4.5 & \textsuperscript{4}I\textsubscript{9/2}            \\
61            & Promethium   & Pm     & {[}Kr{]}4$f$\textsuperscript{4}5$s$\textsuperscript{2}5$p$\textsuperscript{6}  &4  & 2   & 6                       &4   & \textsuperscript{5}I\textsubscript{4}              \\
62            & Samarium     & Sm     & {[}Kr{]}4$f$\textsuperscript{5}5$s$\textsuperscript{2}5$p$\textsuperscript{6}  &5  & 2.5 & 5                       &2.5 & \textsuperscript{6}H\textsubscript{5/2}            \\
63            & Europium     & Eu     & {[}Kr{]}4$f$\textsuperscript{6}5$s$\textsuperscript{2}5$p$\textsuperscript{6}  &6  & 3   & 3                       &0   & \textsuperscript{7}F\textsubscript{0}              \\
64            & Gadolinium   & Gd     & {[}Kr{]}4$f$\textsuperscript{7}5$s$\textsuperscript{2}5$p$\textsuperscript{6}  &7  & 3.5 & 0                       &3.5 & \textsuperscript{8}S\textsubscript{7/2}            \\
65            & Terbium      & Tb     & {[}Kr{]}4$f$\textsuperscript{8}5$s$\textsuperscript{2}5$p$\textsuperscript{6}  &8  & 3   & 3                       &6   & \textsuperscript{7}F\textsubscript{6}              \\
66            & Dysprosium   & Dy     & {[}Kr{]}4$f$\textsuperscript{9}5$s$\textsuperscript{2}5$p$\textsuperscript{6}  &9  & 2.5 & 5                       &7.5 & \textsuperscript{6}H\textsubscript{15/2}           \\
67            & Holmium      & Ho     & {[}Kr{]}4$f$\textsuperscript{10}5$s$\textsuperscript{2}5$p$\textsuperscript{6} &10 & 2   & 6                       &8   & \textsuperscript{5}I\textsubscript{8}              \\
68            & Erbium       & Er     & {[}Kr{]}4$f$\textsuperscript{11}5$s$\textsuperscript{2}5$p$\textsuperscript{6} &11 & 1.5 & 6                       &7.5 & \textsuperscript{4}I\textsubscript{15/2}           \\
69            & Thulium      & Tm     & {[}Kr{]}4$f$\textsuperscript{12}5$s$\textsuperscript{2}5$p$\textsuperscript{6} &12 & 1   & 5                       &6   & \textsuperscript{3}H\textsubscript{6}              \\
70            & Ytterbium    & Yb     & {[}Kr{]}4$f$\textsuperscript{13}5$s$\textsuperscript{2}5$p$\textsuperscript{6} &13 & 0.5 & \multicolumn{1}{l}{3} & 3.5 & \textsuperscript{2}F\textsubscript{7/2}            \\ \hline
\end{tabular}
\end{table}

\begin{figure}[h]
\centering
\includegraphics[width=12cm]{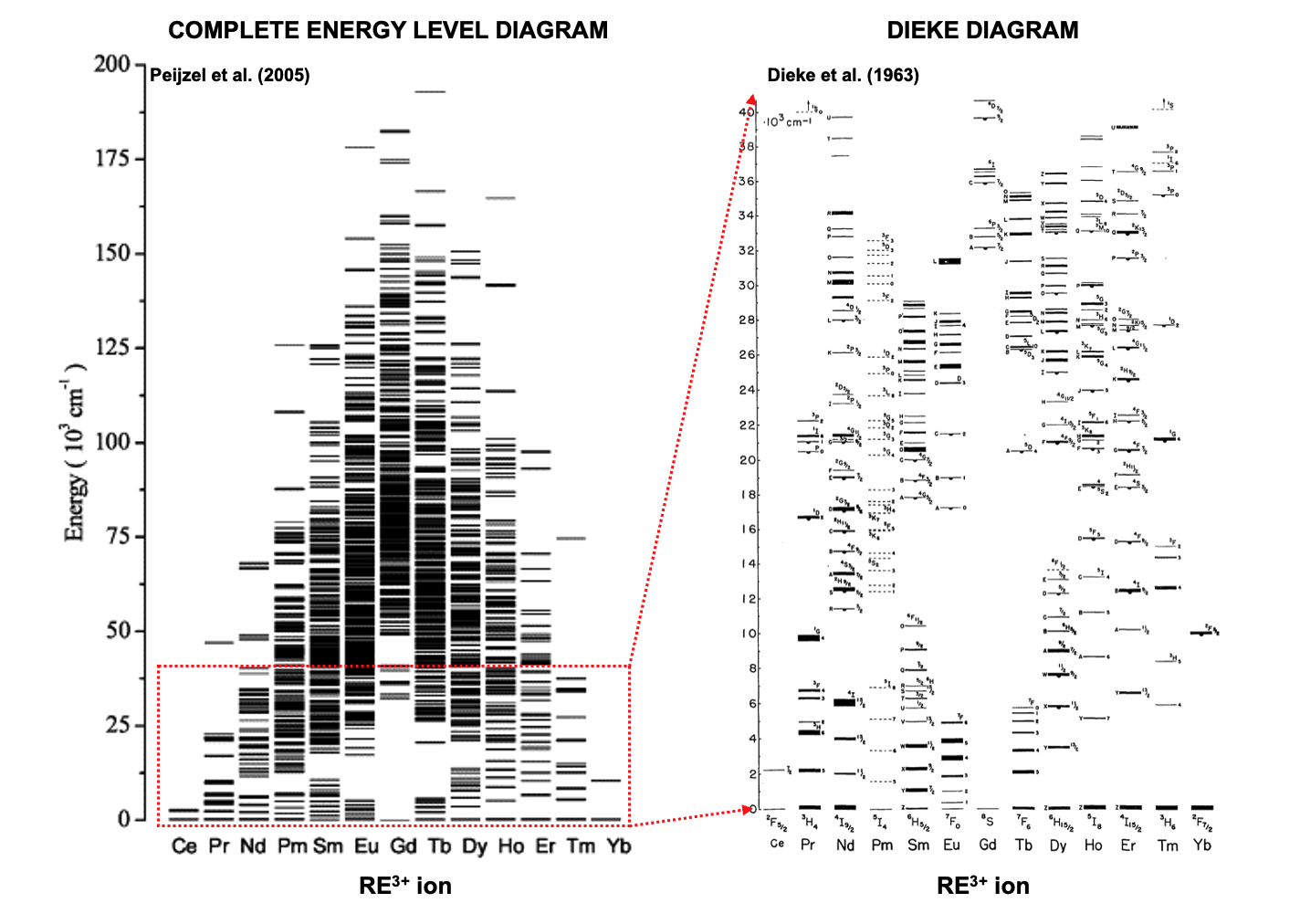}
\caption{\label{fig:loms_energy_dia} Energy level diagram of RE\textsuperscript{3+} ions for the calculated complete set of $^{2S+1}$L$_J$ multiplets\cite{PEIJZEL2005_jo} (left) and the classical experimentally determined "Dieke" \cite{Dieke_1963_jo} diagram for energies up to 40 000 cm\textsuperscript{$-$1}(right).}
\end{figure}

 The electrostatic interaction among electrons then results in the splitting of energy levels by approximately 10\textsuperscript{4} cm\textsuperscript{$-$1}, leading to the formation of new \textsuperscript{2$S$+1}L energy levels separated by the same order of magnitude. Further splitting of these energy levels to new \textsuperscript{2$S$+1}L\textsubscript{$J$} levels occurs when spin-orbit coupling is considered. The influence of ligand field perturbations subsequently generates Stark levels, a process referred to as Stark splitting which divides each $J$ level into 2$J$+1 new Stark levels with energy separation of $\approx$10\textsuperscript{2} cm\textsuperscript{$-$1}. Used spectroscopic symbols describe the total spin angular momentum $S= \sum s\textsubscript{i}$ and total orbital angular momentum  $L= \sum l\textsubscript{i}$ of electron spins $s$\textsubscript{i} and orbital angular momenta $l$\textsubscript{i} for a given electron configuration of RE\textsuperscript{3+} ion. The term symbol \textsuperscript{2$S$+1}L\textsubscript{$J$} of the ground state of a multi-electron atom can be found according to three (1)$-$(3) Hund's rules, where the lowest energy term is that which (1) has the greatest spin multiplicity and (2) the largest value of the total orbital angular momentum (at the maximum multiplicity). Spin-orbit coupling then split \textsuperscript{2S+1}L terms into levels according to the (3) subshell occupancy. If the subshell is less than half full, the lowest energy belongs to the level with the lowest total angular momentum value, $J = |L-S$|,  and on the opposite, if the subshell is exactly or more than half full, the lowest energy belongs to the level with the highest total angular momentum value, $J = |L+S$|. This can be demonstrated on the example of Er\textsuperscript{3+} cation with electron charge configuration of [Xe]4f\textsuperscript{11} with eleven electrons in 4f orbital, where only three are unpaired. By employing the first and second Hund's rule, the total multiplicity is equal to $S = 1.5$ and the largest total orbital angular momentum is equal to $L = 6$. Using the standard notation, the letter symbol of total orbital angular momentum L = S,P,D,F,G,H,I corresponds to $L$ = 0,1,2,3,4,5 and 6. According to the third rule, the subshell is more than half full and thus the total angular momentum value is $J = L+S = 6+1.5 = 15/2$. Described procedure thus results in the construction of \textsuperscript{2\textit{S}+1}L\textsubscript{\textit{J}} ground term for erbium 3+ ion as \textsuperscript{4}I\textsubscript{15/2}. Similar information for other RE ions is listed in Table. \ref{tab:loms_reions}. Extended energy diagram derived from optical experiments by Dieke et al.\cite{Dieke_1963_jo} is presented in Fig.\ref{fig:loms_energy_dia} for the subset of \textsuperscript{2$S$+1}L\textsubscript{$J$} multiplets and energies up to $\approx$5 eV ($\approx$40 000 cm\textsuperscript{$-$1} or $\approx$250 nm). Presented energy levels are placed across the wavelength range covered by commonly used spectroscopic techniques and thus covers only a low-energetic part of the energy level diagram (see Fig.\ref{fig:loms_energy_dia}) for the complete set of   \textsuperscript{2$S$+1}L\textsubscript{$J$} multiplets for each RE\textsuperscript{3+} ion, which was later completed using the theoretical calculations by Peijzel et at.\cite{PEIJZEL2005_jo}

\subsection*{Judd-Ofelt theory}
JO theory was introduced independently to each other by Brian R. Judd\cite{Judd1962} and George S. Ofelt\cite{Ofelt962} in 1962 based on the previous work of J.H. Van  Vleck about spectroscopic properties of rare-earth ions in solids\cite{vleck_1937}. Sharp spectroscopic lines of RE\textsuperscript{3+} ion implicated the intra-4$f$ electronic transitions that occur between the levels inside the 4$f$ electronic shell. This is, however, forbidden by the Laporte selection rule which says that states with even parity can be connected by electric dipole transitions only with states of odd parity and the same in vice versa. Among the other proposed but incorrect explanations based on (1) 4$f$ to 5$d$ transitions or (2) magnetic dipole or electric quadrupole radiation, Van Vleck\cite{vleck_1937} and Broer\cite{BROER_1945_JO} presented a reasonable solution based on the distortion of the electronic motion by surrounding crystal/ligand field in the material. Presented distortions then bypass the Laporte selection rule and allow the electric dipole radiation even for intra-4$f$ electronic transitions.
However, to disturb the wavefunctions and negate the Laporte rule, the external field must also be noncentrosymmetric. From this point, about a quarter of a decade later and with further advances in algebra, computing, and increased applications of lasers, JO\cite{Judd1962,Ofelt962} theory was presented and described the induced electric dipole transitions of RE\textsuperscript{3+} ions in host materials.

JO theory then provides a theoretical expression for the calculation of electric-dipole-induced (ED) oscillator strengths, $f_{ED}^{abs}$ (Eq. \ref{eq:jo_ftheor}), as the ratio between absorbed (emitted) and emitted (absorbed) intensity of electromagnetic radiation for harmonically oscillating electron and expresses the probability of individual J$\leftrightarrow$J' transition as follows,  

\begin{equation}
\label{eq:jo_ftheor}
f_{ED}^{abs}(J \rightarrow J')= \frac{8\pi\textsuperscript{2}m\textsubscript{e}c}{3h\overline\lambda (2J+1)} n\left(\frac{n\textsuperscript{2}+2}{3n}\right)^2 \sum_{i=2,4,6} \Omega_i | \langle (S,L)J || U^{(i)} || J'(S',L') \rangle |\textsuperscript{2},
\end{equation}

\begin{equation}
\label{eq:fmd}
f_{MD}^{abs}(J \rightarrow J')= \frac{h}{6m_ec\lambda} \frac{n}{(2J+1)} | \langle (S,L)J || L+gS || J'(S',L') \rangle |\textsuperscript{2},
\end{equation}

where $J$ and $J'$ are the quantum numbers of the initial ground state and excited state, respectively, $n$ is the refractive index, $h$ is the Planck's constant, $m$\textsubscript{e} is electron mass, $c$ is the speed of light in vacuum, $\overline\lambda$ is the mean wavelength of corresponding \textit{J}$\rightarrow$\textit{J}' transition and $\Omega$\textsubscript{i} are the JO parameters for $i$ = 2, 4, 6. The terms in brackets are the squared reduced matrix elements, which are almost independent on the host matrix. Note, that the summation over $i$ is also known as manifold linestrength which will be introduced later in this section. Compared to the ED-induced absorption, the magnetic-dipole (MD) transitions are usually orders of magnitude smaller. However, some MD transitions can make a significant contribution to the total oscillator strength, $f_{total}^{abs}$. The MD-induced oscillator strength, $f_{MD}^{abs}$, for $J\rightarrow J'$ is then expressed via Eq.\ref{eq:fmd} and unlike ED-induced transitions does not contain any intensity scaling parameter\cite{edgar_fmd_2006,HEHLEN2013_juddofelt}. The reduced matrix element for each transition in Eq. \ref{eq:fmd} is calculated using procedure described in Refs.\cite{wybourne_1965,Walsh2006,HEHLEN2013_juddofelt} and are nonzero only if $S=S'$ and $L=L'$
while $J=J', J=J'+1,$ and $J=J-1$. On the example of Er-doped materials, only the fundamental absorption \textsuperscript{4}I\textsubscript{15/2}$\rightarrow$\textsuperscript{4}I\textsubscript{13/2} ($\approx$1550 nm) contributes significantly. As a result, the magnetic-dipole contribution can account for up to one-third of the total oscillator strength \cite{loms_0}. The total theoretical oscillator strength for transitions which contains both ED and MD is thus given as $f_{total}^{abs}(J \rightarrow J')=f_{ED}^{abs}(J \rightarrow J')+f_{MD}^{abs}(J \rightarrow J')$.
Interaction between the surrounding host matrix and RE\textsuperscript{3+} ions are then expressed by the set of three JO phenomenological parameters, which can be obtained by equating the expressions for the experimental ($f$\textsubscript{exp}) and theoretical ($f_{total}^{abs}$) oscillator strengths using the least-squares method, including both ED+MD or only ED contribution. The experimental oscillator strengths can be calculated from optical absorption spectra using the Eq.\ref{eq:jo_fexp}, 

\begin{equation}
\label{eq:jo_fexp}
f\textsubscript{exp}(J \rightarrow J')= \frac{2m\textsubscript{e}c}{\alpha_f h \overline\lambda^2 N} \int \alpha (\lambda)d\lambda,
\end{equation}

where $\alpha$\textsubscript{f} is fine structure constant, $N$ is rare-earth ion concentration and $\alpha (\lambda)$ is wavelength-dependent absorption coefficient. Optical absorption can be also expressed using the absorption cross section, $\sigma\textsubscript{abs}$, defined as $\sigma\textsubscript{abs}= \alpha (\lambda)/N$. 

Using knowledge of the JO parameters, several important spectroscopic quantities can be calculated for a specific material system, such as the transition probabilities, $A (J',J)$, radiative lifetimes, $\tau\textsuperscript{JO}\textsubscript{r}$, or the luminescence branching ratios, $\beta (J',J)$. The transition probabilities for each transition are calculated from Eq. \ref{eq:jo_transprob}:
\begin{equation}
\label{eq:jo_transprob}
A(J'\rightarrow J)= \frac{64\pi\textsuperscript{4}e^2}{3h\lambda_B^3 (2J'+1)} \left( \chi\textsubscript{ED} S\textsubscript{ED} + \chi\textsubscript{MD}S\textsubscript{MD} \right) ,
\end{equation}

where $J'$ is the total angular momentum of the upper excited state, $e$ is the unit charge of electron, $\lambda_B$ is the transition wavelength (also called Barycenter), $S$\textsubscript{ED} and $S$\textsubscript{MD} are electric and magnetic dipole line strengths and $\chi$\textsubscript{ED} and $\chi$\textsubscript{MD} are the local field corrections of the electric dipole (Eq.\ref{eq:jo_eldip_cor}) and the local field correction of the magnetic dipole (Eq.\ref{eq:jo_magdip_cor}).

\begin{equation}
\label{eq:jo_eldip_cor}
\chi\textsubscript{ED}=n\left( \frac{n\textsuperscript{2}+2}{3} \right)^2 ,
\end{equation}

\begin{equation}
\label{eq:jo_magdip_cor}
\chi\textsubscript{MD}=n^3 ,
\end{equation}

The electric dipole linestrength is then easily calculated from each excited state manifold to lower lying manifold using the JO parameters and reduced matrix elements by Eq. \ref{eq:jo_sed}:
\begin{equation}
\label{eq:jo_sed}
S\textsubscript{ED}= \sum_{i=2,4,6} \Omega_i | \langle (S,L)J || U^{(i)} || J'(S',L') \rangle |\textsuperscript{2},
\end{equation}

\begin{equation}
\label{eq:jo_smd}
S\textsubscript{MD}=\left( \frac{h}{4 \pi m\textsubscript{e}c} \right)^2  | \langle (S,L)J || \hat{L}+g\hat{S} || J'(S',L') \rangle |\textsuperscript{2},
\end{equation}

The magnetic dipole line strengths are given by Eq. \ref{eq:jo_smd}, where \textit{g} is the electron g-factor (g$\approx2.002$) and the terms in brackets are reduced matrix elements of the $||L+gS||$ operator. However, it should be noted that although the expressions for calculating $S_{ED}$ (Eq.~\ref{eq:jo_sed}) and $S_{MD}$ (Eq.~\ref{eq:jo_smd}) are commonly presented in the form shown above, typically using units of cm\textsuperscript{2} and with values on the order of $10^{-20}$, or reported as dimensionless numerical values (especially when tabulated or normalized), the correct physical interpretation of these quantities must correspond to the square of the dipole moment. Consequently, the expressions should include the square of the elementary charge, which is already incorporated in our case in the numerator of the prefactor in Eq.~\ref{eq:jo_transprob}. The correct physical unit is thus that of the squared electric dipole element, namely $\mathrm{C}^2\cdot\mathrm{m}^2$, or equivalently in SI base units, $\mathrm{kg}\cdot\mathrm{m}^4\cdot\mathrm{s}^{-2}$. However, to maintain continuity with the literature and for better clarity, the equations will be used as shown above. The radiative lifetimes of each level, $\tau\textsuperscript{JO} \textsubscript{r}$, are then calculated from the transition probabilities using Eq.\ref{eq:jo_tau_jo}. The luminescence branching ratio, $\beta(J',J)$ is given by Eq.\ref{eq:jo_branching} and represents the distribution of the emission transitions in the emission spectra. Combining the theoretical JO lifetime and branching ratio with the experimentally measured lifetime, $\tau \textsubscript{r}$, for a designated transition results in Eq.\ref{eq:jo_quantum}, which defines the radiative quantum yield, $\eta$, of the corresponding $J' \rightarrow J$ electronic transition.

\begin{equation}
\label{eq:jo_tau_jo}
\tau\textsuperscript{JO} \textsubscript{r}= \frac{1}{\sum_{J'}A(J',J)},
\end{equation}

\begin{equation}
\label{eq:jo_branching}
\beta (J',J)= \frac{A (J',J)}{\sum_{J'}A(J',J)},
\end{equation}

\begin{equation}
\label{eq:jo_quantum}
\eta (J',J)= \frac{\tau\textsubscript{r}}{\tau\textsuperscript{JO}\textsubscript{r}}\beta(J',J),
\end{equation}

\subsection*{Error evaluation of Judd-Ofelt analysis} 

Another important and relevant information is also the reported quality of the performed least-squares fit, which can be quantified  by the RMS parameter, expressed by Eq. \ref{eq:rmsmain_a} or Eq. \ref{eq:rmsmain_b}, respectively:

\begin{subequations}\label{eq:rmsmain}
\begin{align}
RMS_f = \sqrt{\frac{\sum (f\textsubscript{exp}-f_{total}^{abs})^2}{W-3}} \label{eq:rmsmain_a} \\
RMS_S = \sqrt{\frac{\sum (S\textsubscript{exp}-S\textsubscript{total})^2}{W-3}} \label{eq:rmsmain_b}
\end{align}
\end{subequations}

where \textit{W} is the number of transitions used for the calculation. Although RMS values are commonly reported in the literature as indicators of fit quality, the uncertainties of individual Judd–Ofelt parameters are usually not provided, and their calculation is not widely adopted. Here we would like to present the approach based on the error theory of least-square fit described in Ref.\cite{Carnall1968_error1}. The errors of the JO parameters, denoted as $\Delta \Omega_i$ (\textit{i}= 2,4,6), can be thus derived using the oscillator strengths (Eq.\ref{eq:delta_main_a}) or line strengths (Eq.\ref{eq:delta_main_b}), where $f_{ii}$ and $S_{ii}$ are diagonal elements of $3\times 3$ matrix F or S shown in Eq.\ref{eq:mat_main_a} and Eq.\ref{eq:mat_main_b}, respectively.

\begin{subequations}\label{eq:delta_main}
\begin{align}
\Delta \Omega_i = \sqrt{f_{ii}}\textit{RMS}_f
 \label{eq:delta_main_a} \\
\Delta \Omega_i = \sqrt{S_{ii}}\textit{RMS}_S \label{eq:delta_main_b}
\end{align}
\end{subequations}

\begin{subequations}\label{eq:mat_main}
\begin{align}
\text{F} = \begin{pmatrix}
f_{11} & f_{12} & f_{13} \\
f_{21} & f_{22} & f_{23} \\
f_{31} & f_{32} & f_{33}
\end{pmatrix}
 \label{eq:mat_main_a} \\
\text{S} = \begin{pmatrix}
S_{11} & S_{12} & S_{13} \\
S_{21} & S_{22} & S_{23} \\
S_{31} & S_{32} & S_{33}
\end{pmatrix} \label{eq:mat_main_b}
\end{align}
\end{subequations}

The matrix F (or S) is the inverse of the product of the transpose of matrix \textit{u} (for oscillator strength, Eq.\ref{eq:u_main_a}) or \textit{u'} (for line strength, Eq.\ref{eq:u_main_b})
and itself as can be seen in Eq.\ref{eq:product_main_a} or Eq.\ref{eq:product_main_b}, respectively. The matrix element $U_w^{(i)}$ (\textit{i}=2,4,6) is the reduced matrix element of corresponding transition \textit{w}, where \textit{W} is the total number of involved transitions, and $B_w$ is the proportional coefficient (Eq.\ref{eq:b_error}). Note that the used value of the refractive index is connected with the corresponding transition and thus the different used mean transition wavelength, $\overline{\lambda}$. It also should be noted that there might be slight differences in the errors calculated by the two different approaches, using \textit{RMS}\textsubscript{S} or \textit{RMS}\textsubscript{f}, due to the presence of the wavelength-dependent factor $B_w$ in the latter. The topic is described in detail in Ref.\cite{Zhang2020_error2}. In the case of the LOMS.cz online tool, a formula incorporating the line strength is used to estimate the uncertainties of the individual parameters.

\begin{subequations}\label{eq:product_main}
\begin{align}
u = \begin{pmatrix}
B_1 U_1^{(2)} & B_1 U_1^{(4)} & B_1 U_1^{(6)} \\
B_2 U_2^{(2)} & B_2 U_2^{(4)} & B_2 U_2^{(6)} \\
\vdots & \vdots & \vdots \\
B_w U_w^{(2)} & B_w U_w^{(4)} & B_w U_w^{(6)}
\end{pmatrix}
 \label{eq:u_main_a} \\
u' = \begin{pmatrix}
U_1^{(2)} & U_1^{(4)} & U_1^{(6)} \\
U_2^{(2)} & U_2^{(4)} & U_2^{(6)} \\
\vdots & \vdots & \vdots \\
U_w^{(2)} & U_w^{(4)} & U_w^{(6)}
\end{pmatrix} \label{eq:u_main_b}
\end{align}
\end{subequations}

\begin{subequations}\label{eq:product_main}
\begin{align}
\text{F} = (u^Tu)^{-1} = 
\begin{pmatrix}
\displaystyle \sum_{w=1}^{W} \left( B_w U_w^{(2)} \right)^2 &
\displaystyle \sum_{w=1}^{W} B_w^2 U_w^{(2)} U_w^{(4)} &
\displaystyle \sum_{w=1}^{W} B_w^2 U_w^{(2)} U_w^{(6)} \\[10pt]
\displaystyle \sum_{w=1}^{W} B_w^2 U_w^{(2)} U_w^{(4)} &
\displaystyle \sum_{w=1}^{W} \left( B_w U_w^{(4)} \right)^2 &
\displaystyle \sum_{w=1}^{W} B_w^2 U_w^{(4)} U_w^{(6)} \\[10pt]
\displaystyle \sum_{w=1}^{W} B_w^2 U_w^{(2)} U_w^{(6)} &
\displaystyle \sum_{w=1}^{W} B_w^2 U_w^{(4)} U_w^{(6)} &
\displaystyle \sum_{w=1}^{W} \left( B_w U_w^{(6)} \right)^2
\end{pmatrix}^{-1}
 \label{eq:product_main_a} \\
\text{S} = (u'^Tu')^{-1}=\begin{pmatrix}
\displaystyle \sum_{t=1}^{W} \left( U_w^{(2)} \right)^2 &
\displaystyle \sum_{w=1}^{W} U_w^{(2)} U_w^{(4)} &
\displaystyle \sum_{w=1}^{W} U_w^{(2)} U_w^{(6)} \\[10pt]
\displaystyle \sum_{w=1}^{W} U_w^{(2)} U_w^{(4)} &
\displaystyle \sum_{w=1}^{W} \left( U_w^{(4)} \right)^2 &
\displaystyle \sum_{w=1}^{W} U_w^{(4)} U_w^{(6)} \\[10pt]
\displaystyle \sum_{w=1}^{W} U_w^{(2)} U_w^{(6)} &
\displaystyle \sum_{w=1}^{W} U_w^{(4)} U_w^{(6)} &
\displaystyle \sum_{w=1}^{W} \left( U_w^{(6)} \right)^2
\end{pmatrix}^{-1} \label{eq:product_main_b}
\end{align}
\end{subequations}

\begin{equation}
\label{eq:b_error}
B_w =  \frac{8 \pi^2 m_e c  \left( n^2 + 2 \right)^2 }
{3  h  \overline{\lambda}  (2J + 1)  9  n}
\end{equation}

\subsection*{Judd-Ofelt theory: Experimental practice}
From the experimental perspective, accurate spectroscopic characterization of the prepared materials is essential for the proper application of the JO theory and estimation of JO parameters, transition probabilities and derived values of branching ratios and theoretical luminescence lifetimes. 

\begin{figure}[h]
\centering
\includegraphics[width=13cm]{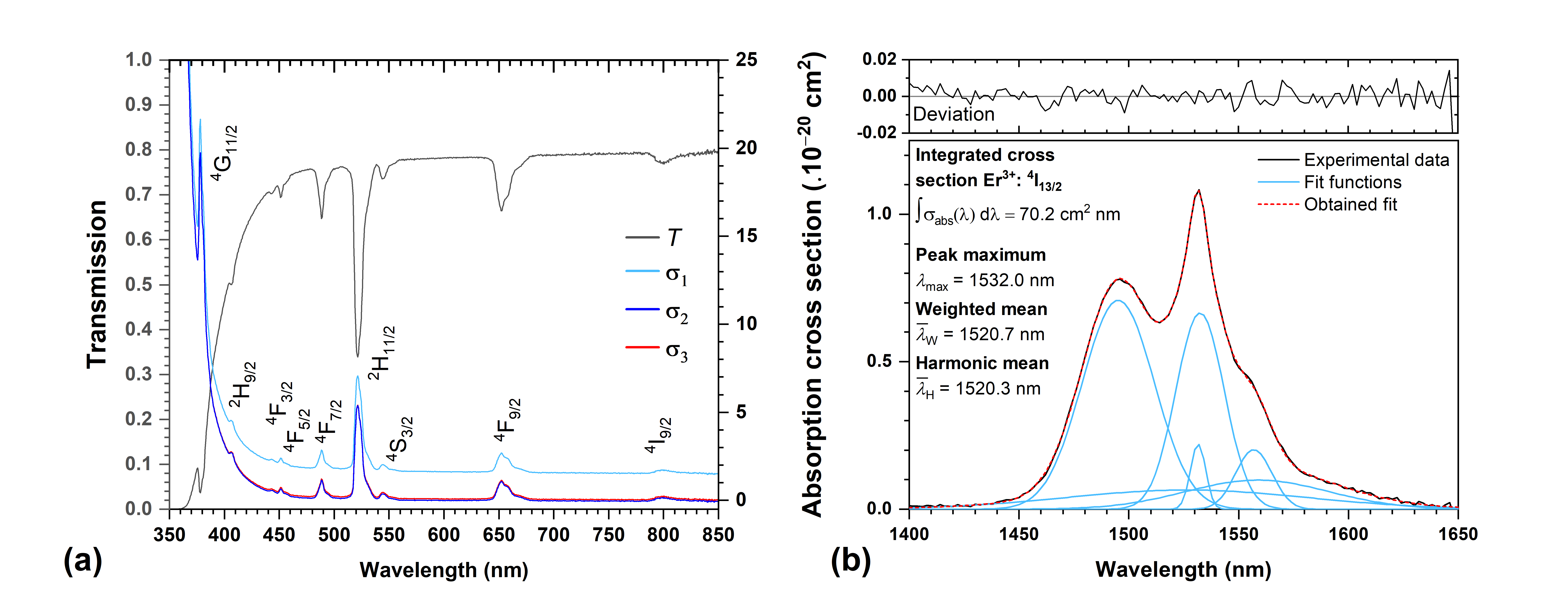}
\caption{\label{fig:jo_porov_sigm} a) Transmission spectrum and corresponding absorption cross sections, employing various corrections on scattering losses or plane parallel geometry of the sample; b) example of integrated area calculation of a selected band.}
\end{figure}

The first step of the JO analysis requires the measurement of the transmission spectrum, $T(\lambda)$, to determine the wavelength-dependent values of the absorption coefficient, $\alpha (\lambda)$, and then the values of the absorption cross-section, $\sigma$\textsubscript{abs}$(\lambda)$. Although the calculation of the $\sigma$\textsubscript{abs}$(\lambda)$ value from the absorption coefficient using the known RE\textsuperscript{3+} ion concentration (\textit{N}) is relatively simple, where $\sigma\textsubscript{abs}( \lambda ) = \alpha\textsubscript{k} ( \lambda ) /N$, the calculation of the absorption coefficient may vary across the literature depending on whether scattering losses are not included (\ref{eq:jo_alpha1}), included (\ref{eq:jo_alpha2}) and if taking into account multiple reflections in plane parallel geometry of the sample (\ref{eq:jo_alpha3}) (in the case of solids). As is visible from Fig.\ref{fig:jo_porov_sigm} in the example of Er\textsuperscript{3+}-doped glass, the spectral shape of corresponding transitions in the transparent region is practically identical with significant offset caused by the not included/included reflectivity (\textit{R}). In cases where the absorption band is offset from the zero $\sigma\textsubscript{abs}( \lambda )$ value or overlaps with the absorption edge, it is therefore necessary to subtract the background to obtain the most possible accurate value. If the number of observed manifolds is sufficient, it is recommended to exclude the transitions within the absorption edge from the calculation of the JO parameters to increase fit accuracy.

\begin{equation}
\label{eq:jo_alpha1}
\alpha\textsubscript{1} = \frac{-1}{l}ln(T) = \frac{-2.303log\textsubscript{10}(T)}{l}
\end{equation}

\begin{equation}
\label{eq:jo_alpha2}
\alpha\textsubscript{2} = \frac{-1}{l} ln \left(\frac{T}{(1-R)^2} \right)=\frac{ 2.303\left[-log\textsubscript{10}(T)+log\textsubscript{10}(1-R)^2\right]}{l}     
\end{equation}

\begin{equation}
\label{eq:jo_alpha3}
\alpha\textsubscript{3} = \frac{1}{l}ln \left[ \frac{ (1-R)\textsuperscript{2}+\sqrt{(1-R)\textsuperscript{4}+4R^2T^2}}{2T}    \right]
\end{equation}

Derived spectral dependence of the $\sigma\textsubscript{abs}(\lambda)$ is used for estimation of the integrated absorption cross section, $\int_{J \rightarrow J'} \sigma\textsubscript{abs}(\lambda)d\lambda$ (in cm\textsuperscript{2} nm), for each manifold (Fig.\ref{fig:jo_porov_sigm}b) which is then used for calculation of the experimental oscillator strength (Eq.\ref{eq:jo_fexp}) or experimental linestrength (Eq.\ref{eq:jo_linestr}) according to 

\begin{equation}
\label{eq:jo_linestr}
S\textsubscript{exp}(J \rightarrow J')= \frac{3ch(2J+1)}{8\pi^3e^2\overline\lambda}n\left(\frac{3}{n^2+2} \right)^2\int_{J\rightarrow J'} \sigma (\lambda)d\lambda,
\end{equation}
 where $J$ is the quantum number representing the total angular momentum of the original ground state, found from the \textsuperscript{2$S$+1}L\textsubscript{$J$} term constructed by using the three Hund's rules (see previous section for detailed description). As the linestrength is typically referred  in cm\textsuperscript{2}, the units and input values for other quantities and constants in presented calculations are used as follows: speed of light, $c\approx3\times10^{10}$cm s\textsuperscript{$-$1}, Planck constant, $h=6.626\times10^{-30}$ cm\textsuperscript{2} kg s\textsuperscript{$-$1}, unit charge of electron, $e=1.5189\times10^{-11}$ cm\textsuperscript{3/2} kg\textsuperscript{1/2} s\textsuperscript{$-$1}, fine structure constant, $\alpha=7.297\times10^{-3}\approx1/137$, and electron mass, $m\textsubscript{e}=9.11\times10^{-11}$kg. The last presented parameter, mean wavelength ($\overline\lambda$), can be found as well from the absorption cross section data using the harmonic, $\overline\lambda\textsubscript{H}$ (Eq.\ref{eq:jo_menawavelength}a), or weighted mean value, $\overline\lambda\textsubscript{W}$ (Eq.\ref{eq:jo_menawavelength}b), for each transition as is illustrated in Fig.\ref{fig:jo_porov_sigm}b. Both derived mean values lead to almost similar results, which, however, may differ from the value of simply subtracting absorption band maximum, $\lambda\textsubscript{max}$. It is worth mentioning that the simply derived value of the absorption cross-section maximum is not suitable for accurate JO analysis calculations. Also note, that for the proper calculation of experimental oscillator strength or experimental linestrength, the values of experimentally determined integrated cross section and mean wavelength must be recalculated after subtraction from the graph and used in (cm\textsuperscript{2} cm) and (cm), respectively.

 \begin{equation}
\label{eq:jo_menawavelength}
\text{(a)  }\overline\lambda\textsubscript{H} = \frac{1}{\frac{\sum\lambda\sigma\textsubscript{abs}(\lambda)}{\sum\sigma\textsubscript{abs}(\lambda)}}=\frac{\sum\sigma\textsubscript{abs}(\lambda)}{\sum\lambda\sigma\textsubscript{abs}(\lambda)} \text{    or    } \text{(b) }      \overline\lambda\textsubscript{W} = \frac{\sum\sigma\textsubscript{abs}(\lambda)\lambda}{\sum\sigma\textsubscript{abs}(\lambda)}
\end{equation}

By completing all of the above characteristics, the JO phenomenological parameters, $\Omega\textsubscript{i} (i=2,4,6)$ are determined by fitting the experimental absorption represented by experimental oscillator strength (Eq.\ref{eq:jo_fexp}) or linestrength (Eq.\ref{eq:jo_linestr}) using the least square method to the theoretical ones considering only the electric-dipole contribution ($f_{total}^{abs}=f_{ED}^{abs}$ or $S_{total}=S_{ED}^{abs}$) or both electric/magnetic-dipole contributions ($f_{total}^{abs}=f_{ED}^{abs}+f_{MD}^{abs}$ or $S_{total}=S_{ED}+S_{MD}$). On the example of the second case, experimental and theoretical linestrengths are written in their respective matrix forms similarly as described in Ref.\cite{Walsh2006} and the sum of the square difference is minimized. Since the JO theory includes only three parameters, more than three absorption manifolds have to be provided for calculation, and thus JO theory cannot be applied to single Yb\textsuperscript{3+}-doped materials. After fitting procedure, materials characteristics, such as $A(J',J), \beta (J',J)$ and $\tau\textsuperscript{JO}\textsubscript{r}$, are calculated using the known JO parameters from Eq.\ref{eq:jo_transprob}, Eq.\ref{eq:jo_tau_jo} and Eq.\ref{eq:jo_branching}. Nevertheless, for proper calculation of transition probabilities (Eq.\ref{eq:jo_transprob}), it is also necessary to know the value of corresponding transition ($J'\rightarrow J$) wavelength  from the excited state to the ground/lower-energy state, $\lambda_B$, also commonly referred as Barycenter. This value should be in principle different from the mean wavelength $\overline\lambda$ or absorption band maximum ($\lambda\textsubscript{max}$). However, the assignment of the barycenter varies considerably within the literature (or is not clearly explained)\cite{Walsh2006,HEHLEN2013_juddofelt} and can be divided into three main approaches, using the (1) similar value of mean wavelength $\overline\lambda$ derived from the optical absorption measurements as Barycenter or (2) tabulated values assigned with U\textsuperscript{(2)},U\textsuperscript{(4)},U\textsuperscript{(6)} elements regardless of the host material or (3) the peak/mean wavelength derived from emission spectra at room temperature. The first approach is consistent with the original Judd–Ofelt theory postulated by Judd and Ofelt, where time-reversal symmetry is assumed and the Einstein coefficients are based on the same wavelength for both absorption and emission. In this context, the line strength is invariant with respect to the direction of the transition. However, in real materials, non-radiative effects, phonon coupling, or structural relaxation may play a significant role and cause broadening of energy levels, leading to a shift of the emitted light to longer wavelengths. Using the last approach, it is possible to estimate the spectral shift (Stokes shift) between mean absorption and emission wavelength for one transition and then apply this difference to all other transitions. Given the extensive nature of the topic, it is up to the author which approach is chosen and which would best fit the experimental results. Nevertheless, based on theoretical justification and consultations with experts in the field, the authors of this present study recommend using the first approach. This method aligns with the assumptions of the original JO framework and is widely adopted in the literature. To ensure flexibility and enable compatibility with previously published data and alternative methodologies, the LOMS.cz tool incorporates all three options described. A detailed step-by-step implementation for each approach is provided in the subsequent sections.

\subsection*{Combinatorial Judd-Ofelt theory}
Following the previous section, it is clear that the selection of the appropriate transition bands, their experimental description or the decision whether to take into account the magnetic-dipole correction are crucial for accurate calculation of the JO parameters\cite{loms_0,Walsh2006,HEHLEN2013_juddofelt}. Judd-Ofelt analysis then minimizes the square of the difference between theoretical ($f_{total}^{abs}$ or $S_{total}$) and experimentally obtained ($f_{exp}$ or $S_{exp}$) oscillator strengths/linestrengths in the form described above and using the corresponding $\Omega_i (i=2,4,6)$ as an adjustable parameters. To compute the JO parameters, at least four experimentally measured absorption manifolds must be used. When a larger set of measured absorption bands is available, it becomes possible to exclude certain transitions (e.g., those exhibiting hypersensitivity) or to limit the JO analysis to transitions within a specific spectral region, for example, due to experimental limitations or the presence of fundamental absorption of the host matrix. However, for accurate determination of the complete set of all three JO parameters, the following criteria must be met: (1) the involved transitions must have non-zero values of the corresponding reduced squared matrix elements $U^{(i)}$ ($i = 2, 4, 6$), (2) these values should be of the same order of magnitude, and (3) at least three transitions that satisfy the previous two conditions must be used. 

As a result, various studies exclude hypersensitive transitions, such as the ${}^4\text{G}_{11/2} \rightarrow {}^4\text{I}_{15/2}$ transition for Er\textsuperscript{3+} ions with a high $U^{(2)}$ value\cite{Afef_2018_hyper1,Karthikeyan_2015_hyper2,Mariyappan_2019_hyper3}, do not cover the full spectral range due to the lack of experimental capability to measure absorption bands in the NIR/MIR regions (Nd\textsuperscript{3+} (\textsuperscript{4}I\textsubscript{11/2}), Dy\textsuperscript{3+} (\textsuperscript{6}H\textsubscript{13/2}), Sm\textsuperscript{3+} (\textsuperscript{6}H\textsubscript{7/2} and \textsuperscript{6}H\textsubscript{9/2}), etc), include/exclude the transitions with magnetic-dipole contribution or selectively include/exclude transitions affected by the absorption edge. This last scenario can be particularly limiting for materials with low optical transmission in the visible spectral region, such as chalcogenide glasses\cite{Aoki_2017_trans0,Bruno_Bureau_2009_trans1,Seddon1995_trans2}, since this region typically contains the majority of experimentally observable absorption bands associated with rare-earth ions. For some materials, it is therefore in principle necessary to include the transitions affected by the absorption edge, otherwise they would not meet the condition for the minimum number of used manifolds. Using the recently developed Combinatorial Judd-Ofelt analysis (C-JO)\cite{loms_0} and a higher than minimum number of transitions, it is thus possible to identify those manifold combinations that enable accurate JO analysis ensuring consistent and reliable results. Moreover, by employing various types of host materials and broad-spectrum analysis for each eare-earth ion it will be possible to identify such critical combinations, which are essential for the calculation of JO parameters and thus should not be omitted.
The total value of all possible combinations then depends on the number of input absorption bands ($N$\textsubscript{B}) according to Eq.\ref{eq:jo_total_combin}

 \begin{equation}
\label{eq:jo_total_combin}
\text{Total combinations} = \sum_{r=k}^{N\textsubscript{B}} \binom{N\textsubscript{B}}{r} 
\end{equation}

where $k$ is the minimum number of elements in each combination (from 4 to N\textsubscript{B}) and $\binom{N\textsubscript{B}}{r} $ is the binomial coefficient calculated as $\binom{N\textsubscript{B}}{r} = \frac{N\textsubscript{B}!}{r!(N\textsubscript{B}-r)!}$. It is then possible to obtain 5, 22, 64, 163, 382 and 848 possible combinations for original sets composed of 5, 6, 7, 8, 9 and 10 experimentally obtained absorption bands. The obtained set of all possible combinations can be subsequently reduced by inappropriate combinations using different empirical approaches (e.g. due to unphysicality of partial solutions or non-converging results when calculating JO parameters) or using the analysis of statistical distribution of the resulting JO parameter values depending on the absorption bands used as is, for example, described in Ref.\cite{loms_0}.  In order to eliminate the empirical selection approach, the box/whisker plot statistical method may be applied to the original set of all possible combinations reduced by the non-physical cases (negative values of JO parameters)\cite{loms_0}. According to the used statistical model, data points (combinations) outside the whisker boundaries are identified as outliers and thus may be excluded from the dataset as was shown in Ref.\cite{loms_0} on the example of Er\textsuperscript{3+}-doped materials. Several other examples of presented C-JO analysis are given in the following section \textit{Computational Validation:Judd-ofelt analysis and Combinatorial Judd-Ofelt analysis}.

\begin{figure}[h]
\centering
\includegraphics[width=14cm]{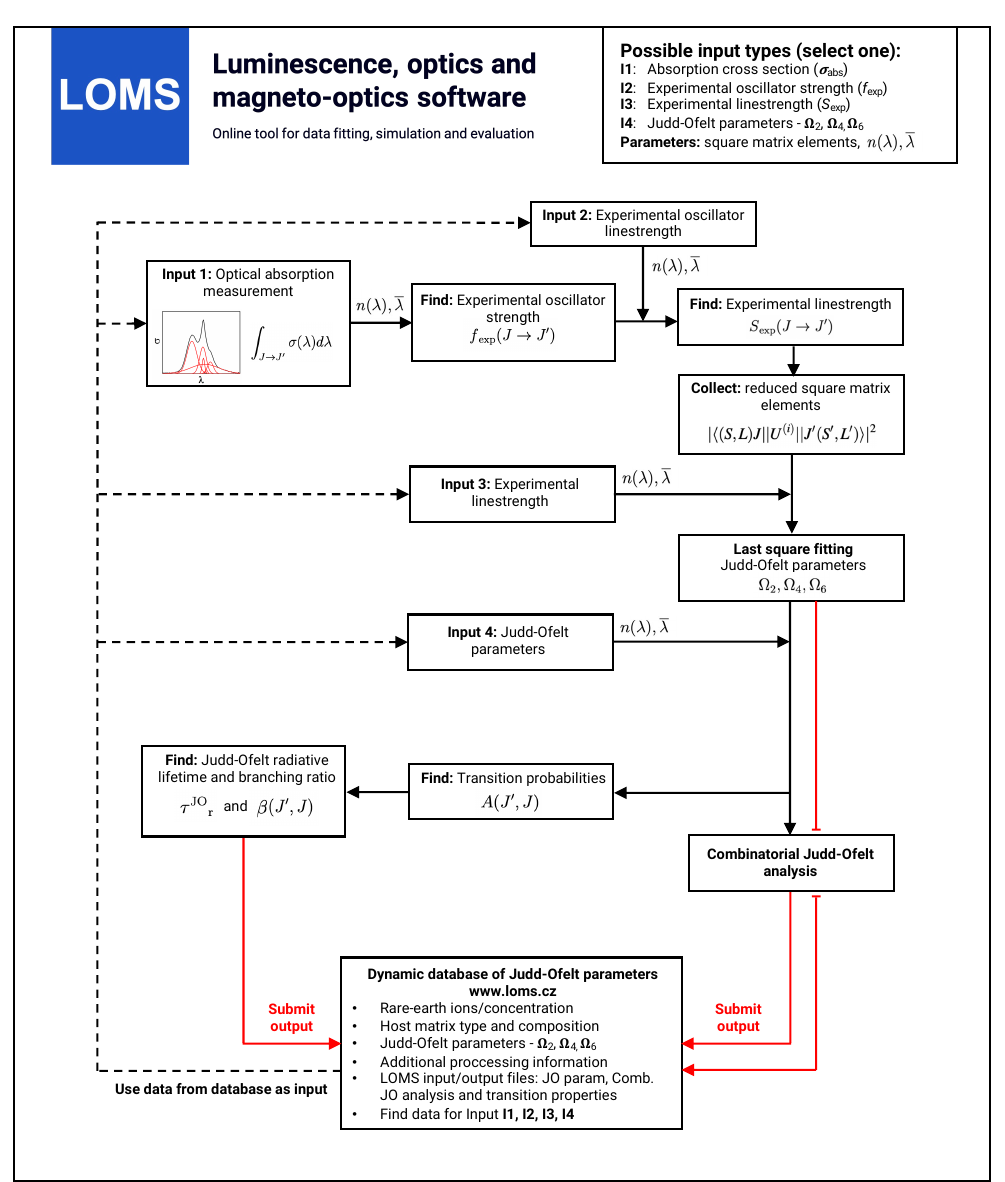}
\caption{\label{fig:jo_flowchart} Software procedure of Judd-Ofelt analysis and implementation of Judd-Ofelt parameters database.}
\end{figure}

\clearpage

\section*{Evaluation protocol and graphical software interface}
\begin{figure}[h]
\centering
\includegraphics[width=14cm]{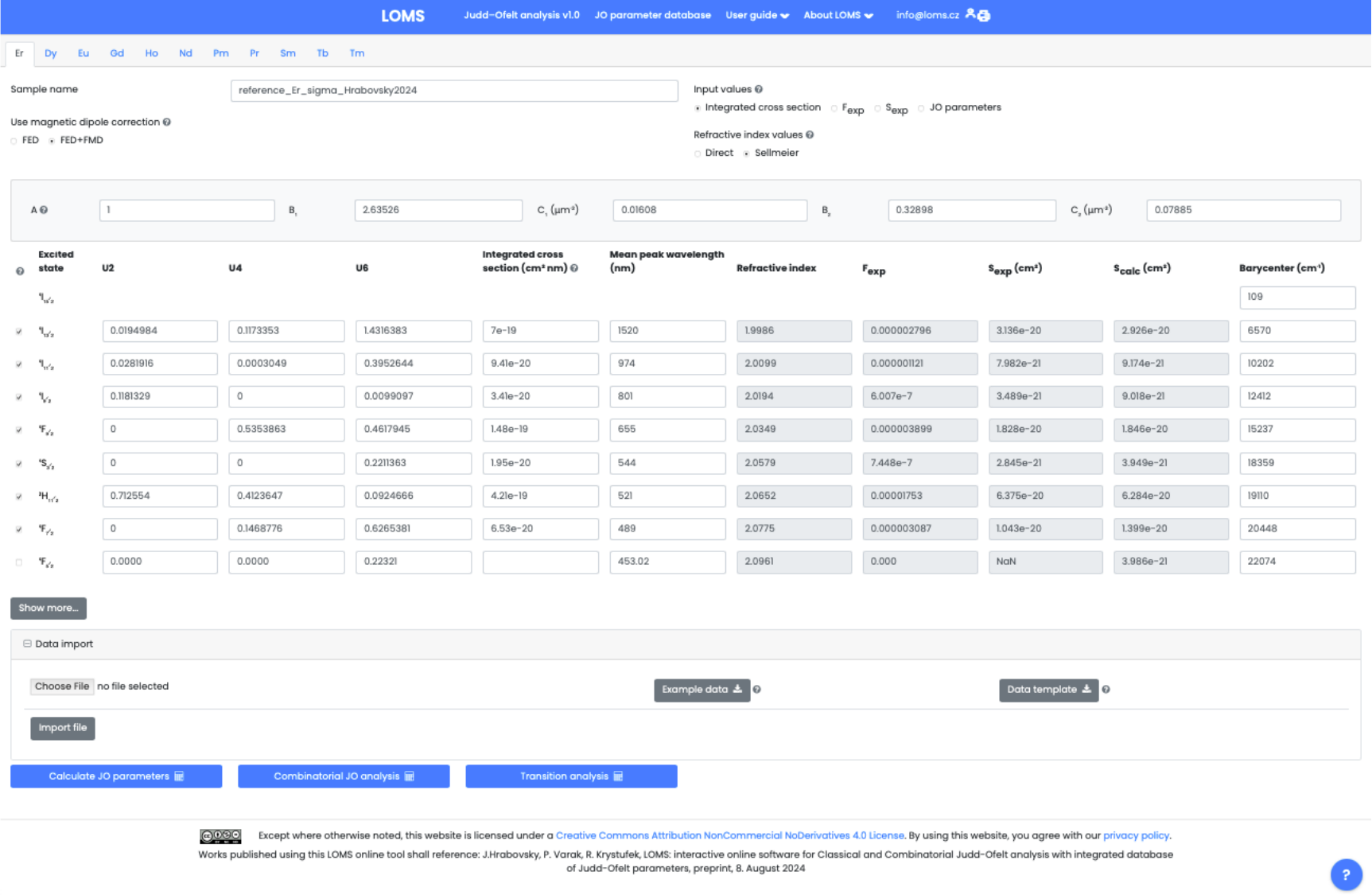}
\caption{\label{fig:jo_gui} The graphical user interface of LOMS online tool, which is available at https://www.LOMS.cz/.}
\end{figure}

The process of JO and Combinatorial JO analysis using the Luminescence, optics and magneto-optics software (LOMS) (www.LOMS.cz or www.LOMS.cz/jo) is outlined in the attached flowchart (Fig.\ref{fig:jo_flowchart}), while the graphical user interface of LOMS computational tool is shown in Fig.\ref{fig:jo_gui}. To enhance versatility, users can choose from four recommended input options, depending on the desired level of data processing and verification (Fig.\ref{fig:jo_gui}, Radio button: \textit{Input values}). The software supports direct processing of experimental oscillator strength/linestrength input data, enabling straightforward comparison with experimental results from the literature. Additionally, a magnetic-dipole correction feature is available for input data in the form of absorption cross sections, which can be applied by selecting Radio button: \textit{Use magnetic dipole correction}. Furthermore, the software allows  for direct input of JO parameters, followed by the calculation of material radiative characteristics. The list of possible input files is as follows:
\begin{enumerate}
  \item Integrated absorption cross section $\int\sigma\textsubscript{abs} d\lambda$ (in cm\textsuperscript{2} nm) or
  \item Experimental oscillator strength, $f$\textsubscript{exp}, taken from an external source or calculated using Eq.\ref{eq:jo_fexp} or
  \item Experimental linestrength, $S$\textsubscript{exp} (in cm\textsuperscript{2}), taken from an external source or calculated using Eq.\ref{eq:jo_linestr} or
    \item Judd-Ofelt parameters, $\Omega_2, \Omega_4, \Omega_6,$ (in cm\textsuperscript{2}), taken from an external source or calculated using aforementioned procedure.
\end{enumerate}

Furthermore, to successfully calculate JO parameters and radiation material characteristics (transition probabilities, radiative lifetimes and branching ratios), the input file must be supplemented with the following data sets for each experimentally derived manifold:

\begin{enumerate}
  \item Refractive index (Fig.\ref{fig:jo_gui}, Radio button: \textit{Refractive index values}) and
  \item Mean peak wavelength, $\overline{\lambda}$, (in nm) derived using Eq.\ref{eq:jo_menawavelength} for each placed transition (Fig.\ref{fig:jo_gui}, Text field: \textit{Mean peak wavelength})
  
  \item Reduced squared matrix elements U\textsuperscript{(2)}, U\textsuperscript{(4)}, U\textsuperscript{(6)} for each placed transition (Fig.\ref{fig:jo_gui}, Text fields: \textit{U2, U4, U6})
    \item Barycenter: (in cm\textsuperscript{$-$1}) for each transition (Fig.\ref{fig:jo_gui}, Text fields: \textit{Barycenter}). If they are not experimentally detectable, it is necessary to use their tabulated values or choose one of the approaches discussed in section: \textit{Judd-Ofelt theory: Experimental practice}. Note, the authors of this study recommends to use similar values as the mean peak wavelength, $\overline{\lambda}$, derived from optical transmission/absorption spectra and thus $\lambda_B =\overline{\lambda_H} \text{ or } \overline{\lambda_W} $ in Eq.\ref{eq:jo_transprob} . For transitions which are out of the measured interval use the tabulated values, for example those involved in example/template files provided in the LOMS.cz software.
\end{enumerate}

\begin{figure}[h]
\centering
\includegraphics[width=10.5cm]{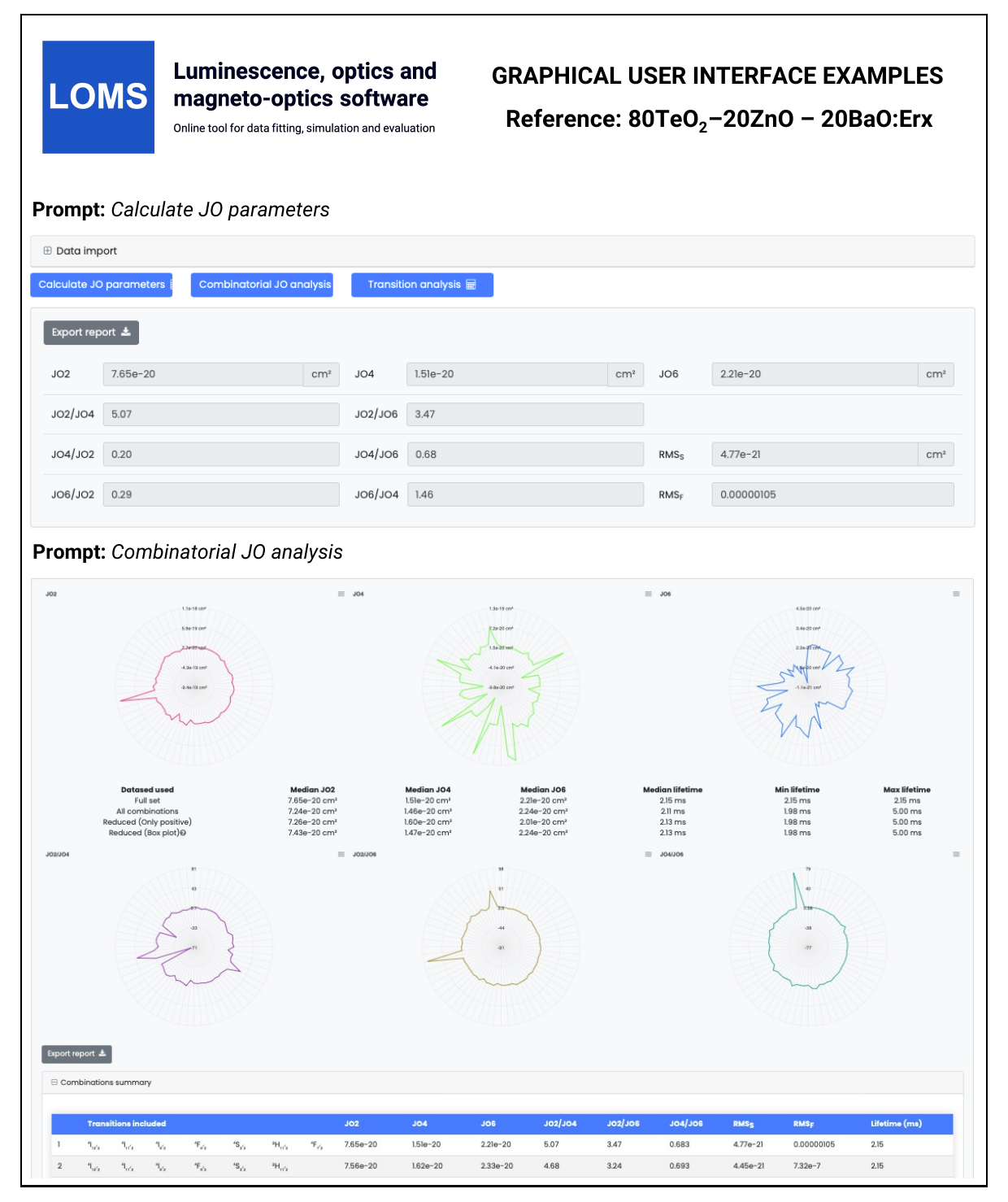}
\caption{\label{fig:loms_gui} The graphical user interface of LOMS online tool (https://www.LOMS.cz): Illustrative example of results structure for classical and combinatorial Judd-Ofelt analysis.}
\end{figure}

The refractive index can be added directly as defined values for each transition in the same row (Fig.\ref{fig:jo_gui}, Text field: \textit{Refractive index}) or expressed using a standard two-term Sellmeir model (Eq.\ref{eq:jo_selm_optika})

\begin{equation}
\label{eq:jo_selm_optika}
n\textsuperscript{2} = A+ \frac{ B\textsubscript{1}\lambda\textsuperscript{2}}{\lambda\textsuperscript{2}-C\textsubscript{1}} + \frac{ B\textsubscript{2}\lambda\textsuperscript{2}}{\lambda\textsuperscript{2}-C\textsubscript{2}},
\end{equation}
where the $A$, $B$\textsubscript{1}, $C$\textsubscript{1}, $B$\textsubscript{2} and $C$\textsubscript{2} are the Sellmeier coefficients. Note, that while refractive index values can be entered directly - sufficient for calculation of JO parameters - determining the radiative characteristics, such as $A(J',J), \beta (J',J)$ and $\tau\textsuperscript{JO}\textsubscript{r}$, requires specifying its spectral dependence via the appropriate Sellmeier coefficients. If the refractive index of the material is not readily available, it can be sourced from publicly accessible databases, such as \textit{refractiveindex.info}\cite{refr_index_info}. A consistent set of tabulated reduced matrix elements for all RE elements listed in Table \ref{tab:loms_reions} and default values of barycenters and mean peak wavelengths are provided (see Figshare repository\cite{redeposit} or www.LOMS.cz) with the possibility of their interactive editing in the software GUI if necessary. A key feature of the software is the ability to dynamically select the number of included transitions — via a column of checkboxes on the left side in Fig.\ref{fig:jo_gui} - without requiring modifications to the input data structure. Once all the above requirements have been met, the classical JO analysis can be performed via pressing Action button: \textit{Calculate JO parameters}, while a Combinatorial JO analysis - evaluating all possible combinations of inserted absorption bands -  can be executed using the Action button: \textit{Combinatorial JO analysis} button. The GUI structure displaying the results is shown in Fig.\ref{fig:loms_gui}. Displayed 
results of C-JO analysis further contain information regarding the median of JO parameters and min/max values of radiative lifetimes from the lowest energy level for different datasets: (1) Full set: contains all absorption bands, $k=N_B$ in Eq.\ref{eq:jo_total_combin}, (2) All combinations: $k\in\langle 4,N_B \rangle$, (3) Reduced (only positive): set of data without discarded combinations, where at least one JO value is negative and (4) Reduced (Box plot): reduced set of all combinations using the Box plot method similarly as in Ref\cite{loms_0}. Radiative lifetimes are also included for each used combination of absorption bands in data export file together with values of an individual uncertainties for each JO parameter ($\Delta$JO2, $\Delta$JO4, $\Delta$JO6) and qualitative fit indicators of $RMS_f$ or $RMS_S$ .

\clearpage

\begin{figure}[h]
\centering
\includegraphics[width=10cm]{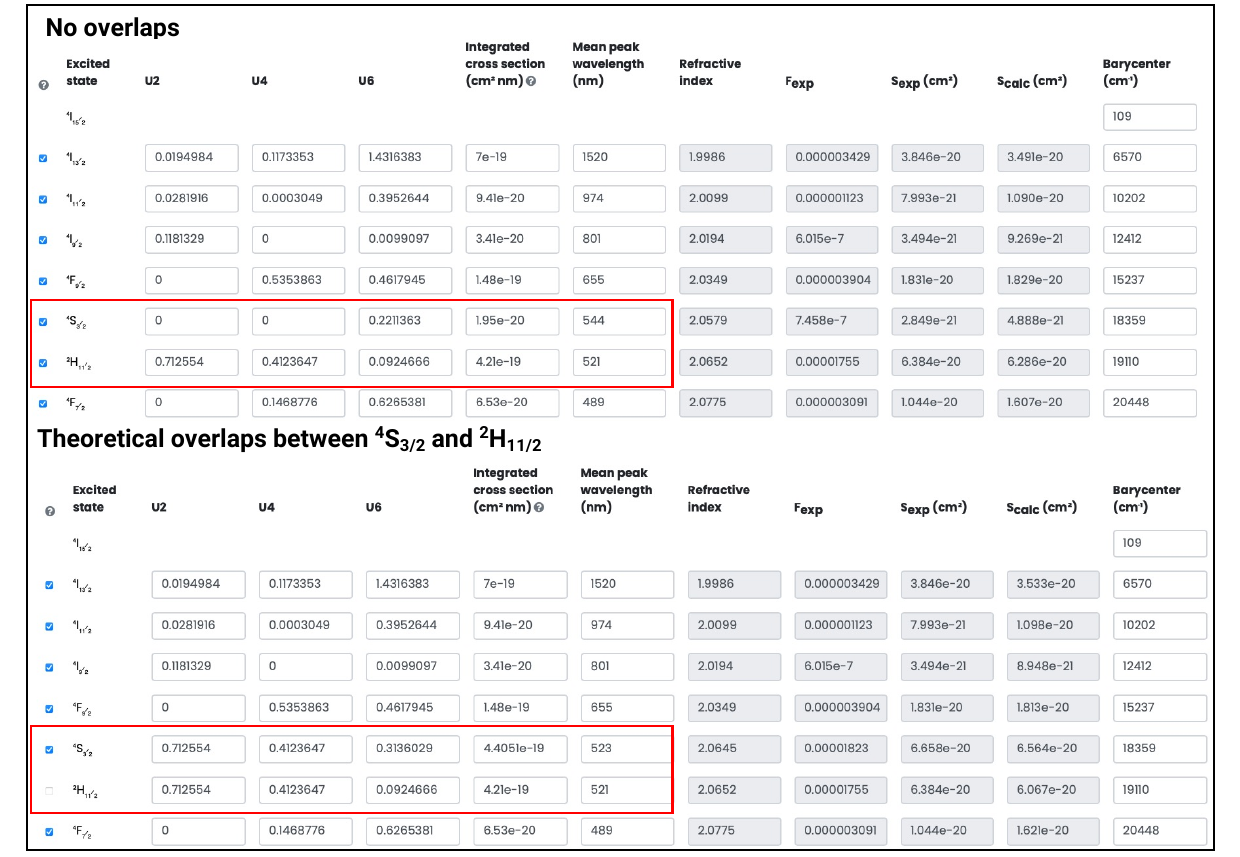}
\caption{\label{fig:loms_gui4_multibands} The graphical user interface of LOMS online tool, with shown comparison between data input structure without and with observed absorption band overlap. See the main text for discussion.}
\end{figure}

Note, that in many cases, two or more closely located transitions may overlap with each other and therefore it is not possible to easily distinguish their independent contribution. This can be the example of two absorption bands \textsuperscript{2}H\textsubscript{11/2} ($\approx$ 530 nm) and \textsuperscript{4}S\textsubscript{3/2} ($\approx$ 550 nm) in Er\textsuperscript{3+}-doped materials. In such cases, it is therefore necessary to apply a modified procedure for the calculation of JO parameters as follows: (1) estimate the combined integrated absorption cross section which involves both absorption bands, (2) estimate the mean peak wavelength in the same way as if it was a single absorption band, (3) sum the respective reduced matrix elements of all the participating transitions into one and (4) write them to the LOMS.cz online GUI in one line - choose the line of one of the involved transitions (or similarly in input .csv file). This modified procedure then affects the U2, U4, U6, integrated cross section and mean wavelength cells. For better clarity, the difference is visible in Fig.(\ref{fig:loms_gui4_multibands}) and the data repository\cite{redeposit} also contains .xls reference file with shown calculation process. It is also important to note, that it is necessary to uncheck the remaining transitions so that only the one combined transition/row participates in the calculation. This then acts as the combined level of \textsuperscript{2}H\textsubscript{11/2} $+$\textsuperscript{4}S\textsubscript{3/2}. It is then necessary to remember that in the output file of the JO analysis and the combinatorial JO analysis, this transition no longer represents only one level, but a combination of all involved manifolds. However, this no longer applies to the calculation of radiative transitions properties ($A, \beta, \tau$), which is done separately and independently of whether the combined or single bands were used for the calculation of JO parameters or not. This is of course due to the fact that radiative properties are calculated directly from the JO parameters, i.e. energy level assignment in \textit{Transition analysis} section is independent of the structure of the data input.

\begin{figure}[h]
\centering
\includegraphics[width=10cm]{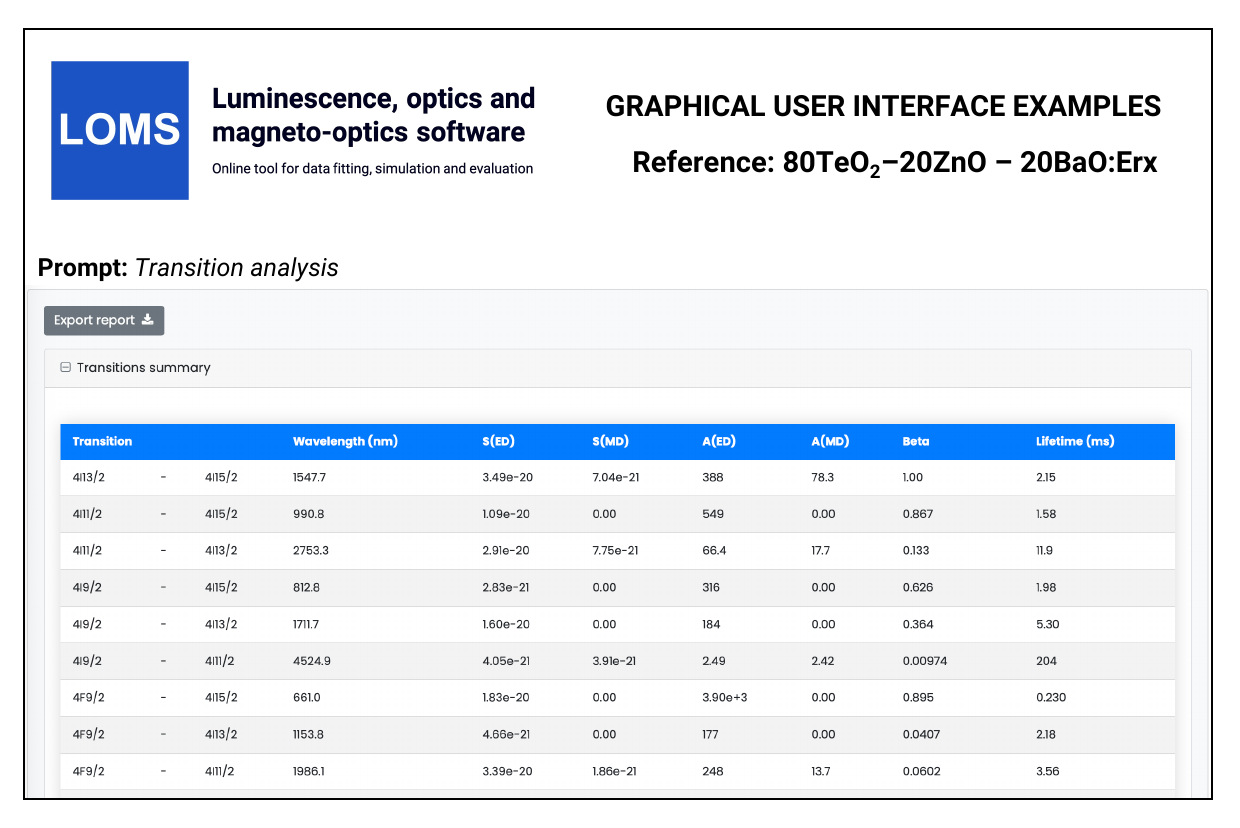}
\caption{\label{fig:loms_gui2} The graphical user interface of LOMS online tool (https://www.LOMS.cz.): Illustrative example of results structure for \textit{Transition analysis}}
\end{figure}

Results of transition analysis, calculation of $A (J',J), \beta(J',J), \tau\textsuperscript{JO}\textsubscript{r}, A\textsubscript{(ED)}, A\textsubscript{MD}$, will be displayed after pressing the Action button: \textit{Transition analysis} button (see Fig.\ref{fig:jo_gui}). The results structure for radiative transition analysis in GUI is shown in Fig.\ref{fig:loms_gui2} and the structure of example output file is visible from Table\ref{tab:trans_er}.  Note, that for successful transition analysis, it is also necessary to include the Barycenter values for each transition and not only for those which were inserted. It is because the transition probabilities, $A (J',J)$ (Eq.\ref{eq:jo_transprob}), are calculated for each transition ($J'\rightarrow$J) from an excited state to the ground/lower-energy state. As was discussed in the section \textit{Judd-Ofelt theory: Experimental practice}, the barycenter value should be in principle different from the mean wavelength $\overline\lambda$ or absorption band maximum ($\lambda\textsubscript{max}$) as the position of photoluminescence emission is usually red-shifted compared to position of optical absorption (this is valid for both peak/mean wavelength values). However, the assignment of the barycenter varies considerably within the literature (or is not clearly explained) and can be divided into three main approaches, using the (1) similar value of mean wavelength $\overline\lambda$ derived from the optical absorption measurements as Barycenter or (2) tabulated values assigned with U\textsuperscript{2},U\textsuperscript{4},U\textsuperscript{6} elements regardless of the host material or (3) the peak/mean wavelength derived from emission spectra at room temperature. To avoid limiting of the calculation, the software allows all the above-mentioned options depending on the selected value. The LOMS.cz software then calculates the energy difference between selected energy levels, which will be used for the calculation of transition probabilities (Eq.\ref{eq:jo_transprob}). The barycenter values may be then inserted as follows:
\begin{enumerate}
  \item \textbf{Barycenter value similar to mean wavelength:} (1) leave the first box for the ground state in Barycenter column (Fig.\ref{fig:jo_gui},Text fields: \textit{Barycenter}) blank or equal to zero, (2) fill the other positions with corresponding recalculated values of mean wavelength in cm\textsuperscript{$-$1} (cm\textsuperscript{$-$1} = $10^{7}$/nm) 
  \item \textbf{Tabulated values of Barycenter:} fill the corresponding manifold cell for each transition using the tabulated values.
  \item \textbf{Barycenter value with the constant shift:} according to software procedure (JOFwin2011) presented by Walsh\cite{Walsh2006}, it is possible to insert the offset position of the ground state which more or less represents the energy spectral shift between optical absorption and emission band peak/mean maximum. In this case, the value in the first box for the ground state in Barycenter column contains the value of this energy spectral shift, whereas the other values represents the mean wavelengths (in cm\textsuperscript{$-$1}) derived from optical absorption spectra. 
\end{enumerate}

Using the last approach, it is possible to estimate the spectral shift between mean absorption and emission wavelength (i.e. Stokes shift) for one transition and then apply this difference to all other transitions. Given the extensive nature of the topic, it is up to the author which approach is chosen and which would best fit the experimental results. However, due to the reasons discussed in section \textit{Judd-Ofelt theory: Experimental practice}, the authors of this study recommend to use the first approach.

\section*{Data Records}
The complete set of blank template input files for each rare-earth ion, illustrative examples of input files together with attached results for JO and C-JO analysis and dataset of JO parameters listed in LOMS.cz database is available at Figshare\cite{redeposit} or the www.LOMS.cz webpage.

It presently, as of February 2025, contains: 
\begin{enumerate}
  \item \textbf{Template files:} complete set of eleven templates for Pr\textsuperscript{3+}, Nd\textsuperscript{3+}, Pm\textsuperscript{3+}, Sm\textsuperscript{3+}, Eu\textsuperscript{3+}, Gd\textsuperscript{3+}, Tb\textsuperscript{3+}, Dy\textsuperscript{3+}, Ho\textsuperscript{3+}, Er\textsuperscript{3+} and Tm \textsuperscript{3+} trivalent rare-earth ions which contains: identification of $J\rightarrow J'$ transition with associated values of reduced matrix elements, mean-wavelenghts and barycenters obtained from Walsh \cite{Walsh2006} JOFwin2011 documentation as a reference.
    \item \textbf{Reference files:} example set of reference files with different types (I.$-$IV. of inputs, Fig.\ref{fig:jo_flowchart}) for JO analysis, C-JO analysis and calculation of radiative properties of Pr\textsuperscript{3+}\cite{loms_87}, Nd\textsuperscript{3+}\cite{loms_93}, Pm\textsuperscript{3+}\cite{loms_46}, Sm\textsuperscript{3+}\cite{loms_266,loms_72}, Tb\textsuperscript{3+}\cite{loms_88}, Dy\textsuperscript{3+}\cite{loms_23,loms_86}, Ho\textsuperscript{3+}\cite{loms_1,loms_90}, Er\textsuperscript{3+}\cite{loms_0,loms_5} and Tm \textsuperscript{3+}\cite{loms_68,loms_45} trivalent rare-earth ions

    \item \textbf{Combinatorial Judd-Ofelt analysis:} output files from C-JO analysis for RE\textsuperscript{3+}-doped materials which contains JO parameters of all possible combinations of involved measured intre-4\textit{f} transitions
    \item \textbf{Database of Judd-Ofelt parameters:} more than 1200 data records of JO parameters and resulting radiative properties for 12 RE\textsuperscript{3+} ions in more than  550 materials/host matrices of various compositions
\end{enumerate}

\subsubsection*{Structure of .csv file import}
\begin{figure}[h]
\centering
\includegraphics[width=10cm]{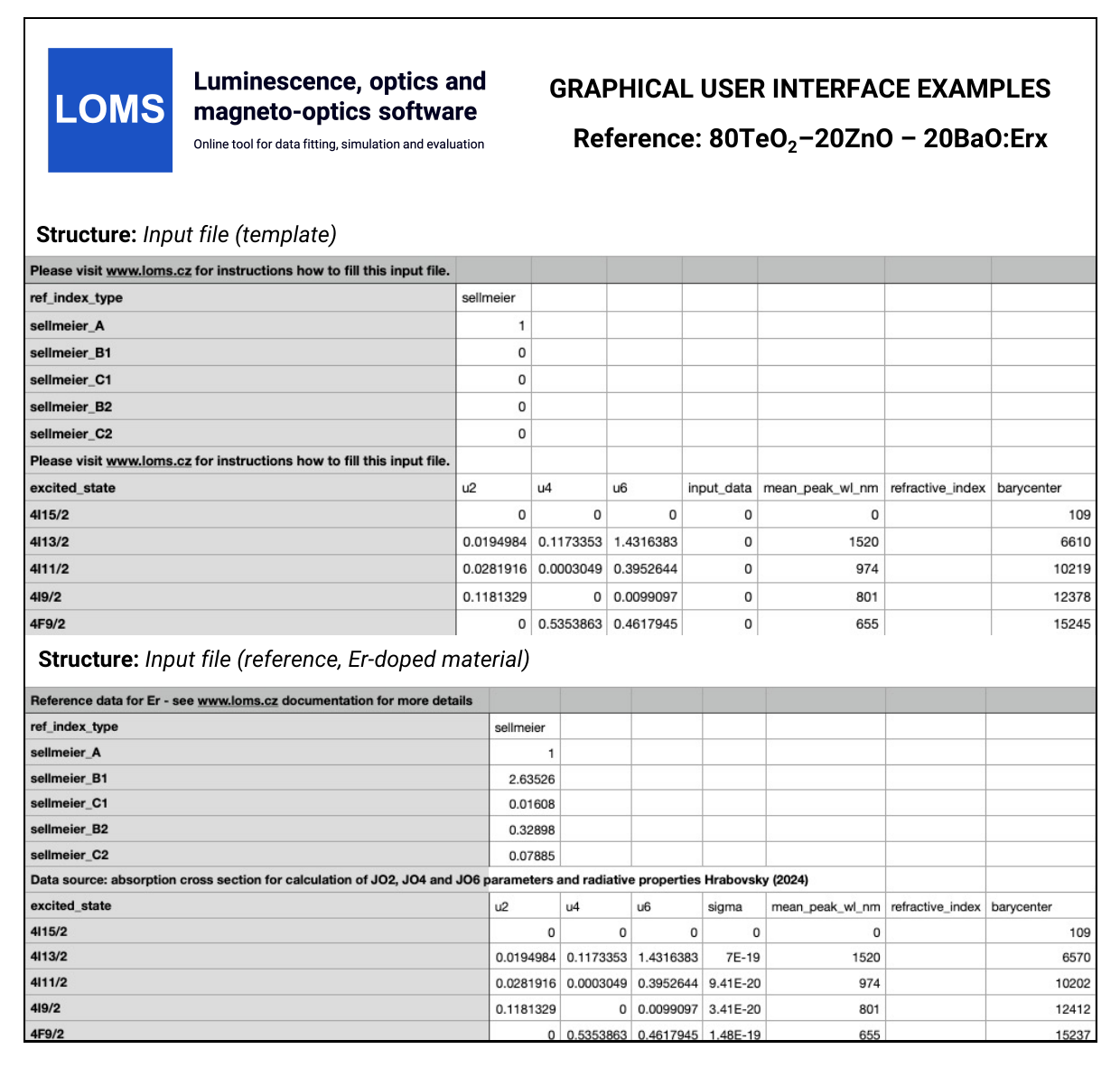}
\caption{\label{fig:loms_gui3_import_exp} The structure of import .csv file.}
\end{figure}

To standardize and simplify the data upload process, users can utilize the option to upload the required data via a .csv file, using the provided templates for all elements. The template .csv file for each RE\textsuperscript{3+} ion is unique and cannot be exchanged between each other since it contains the corresponding absorption transitions notation, assigned reduced squared matrix elements, etc. To successfully complete the form, the following steps must be completed:

\begin{enumerate}
  \item \textbf{Template file:} Download the template file on the relevant rare-earth ion page (www.LOMS.cz/jo) or module documentation (www.LOMS.cz/modules/judd-ofelt-analysis/) or from Figshare data repository\cite{redeposit} 
  \item \textbf{Refractive index import:} Enter the refractive index input structure in the appropriate field following the \textit{ref\_index\_type} cell (see Fig.\ref{fig:loms_gui3_import_exp}) as (1) \textit{sellmeier} for input via Eq.\ref{eq:jo_selm_optika} or (2) \textit{direct} for direct refractive index input. Based on your selection, enter either the Sellmeier coefficients or refractive index values for the corresponding transitions in the column labelled “refractive\_index.” 
  \item \textbf{Transitions and Reduced squared matrix elements:} verify/replace the tabulated Reduced squared matrix elements but do not change the labels of the individual transitions in the first column.
  \item \textbf{Input type:} Select the corresponding form of your input type as follows: absorption cross section (\textit{sigma}), experimental oscillator strength (\textit{fext}), experimental linestrength (\textit{sexp}) or JO parameters (\textit{jo}) and write it down to the cell named \textit{input\_date} (rewrite it). The input values for corresponding transitions have to be placed in the same column. For the selection of JO parameters as an input (only for calculation of radiative properties), please insert the $\Omega_2, \Omega_4, \Omega_6$ JO parameters to the $U\textsuperscript{(2)}, U\textsuperscript{(4)}, U\textsuperscript{(6)}$ of the ground state (replace the zero values).
    \item \textbf{Mean peak wavelength:} Enter the mean peak wavelengths for the transitions for which input data has been provided (see the previous text for proper estimation of mean wavelength value).
    \item \textbf{Barycenter:} Check or provide relevant data for all transitions, otherwise it will not be possible to calculate the relevant radiation characteristics. Please see \textit{Evaluation protocol and graphical software interface} section for more details regarding the proper barycenter selection.
    
\end{enumerate}

\subsubsection*{Structure of .csv output file}
Calculated results of JO analysis, Combinatorial JO analysis and radiative transition properties can be exported in the form of .csv files upon clicking on the button Action button: \textit{Export report} in the corresponding section (see Fig.\ref{fig:loms_gui} and Fig.\ref{fig:loms_gui2}). Example output files for the mentioned references are included in Figshare repository\cite{redeposit} and correspond to the data structrures presented in Fig.\ref{fig:loms_gui} and Fig.\ref{fig:loms_gui2}. Data export from classical JO analysis also contains all input information for selected bands and both experimental and theoretical values of linestrength accompanied by the estimated ratios between calculated JO parameters.

\subsection*{Technical Validation: Judd-Ofelt analysis and Combinatorial Judd-Ofelt analysis}
\begin{figure}[h]
\centering
\includegraphics[width=13cm]{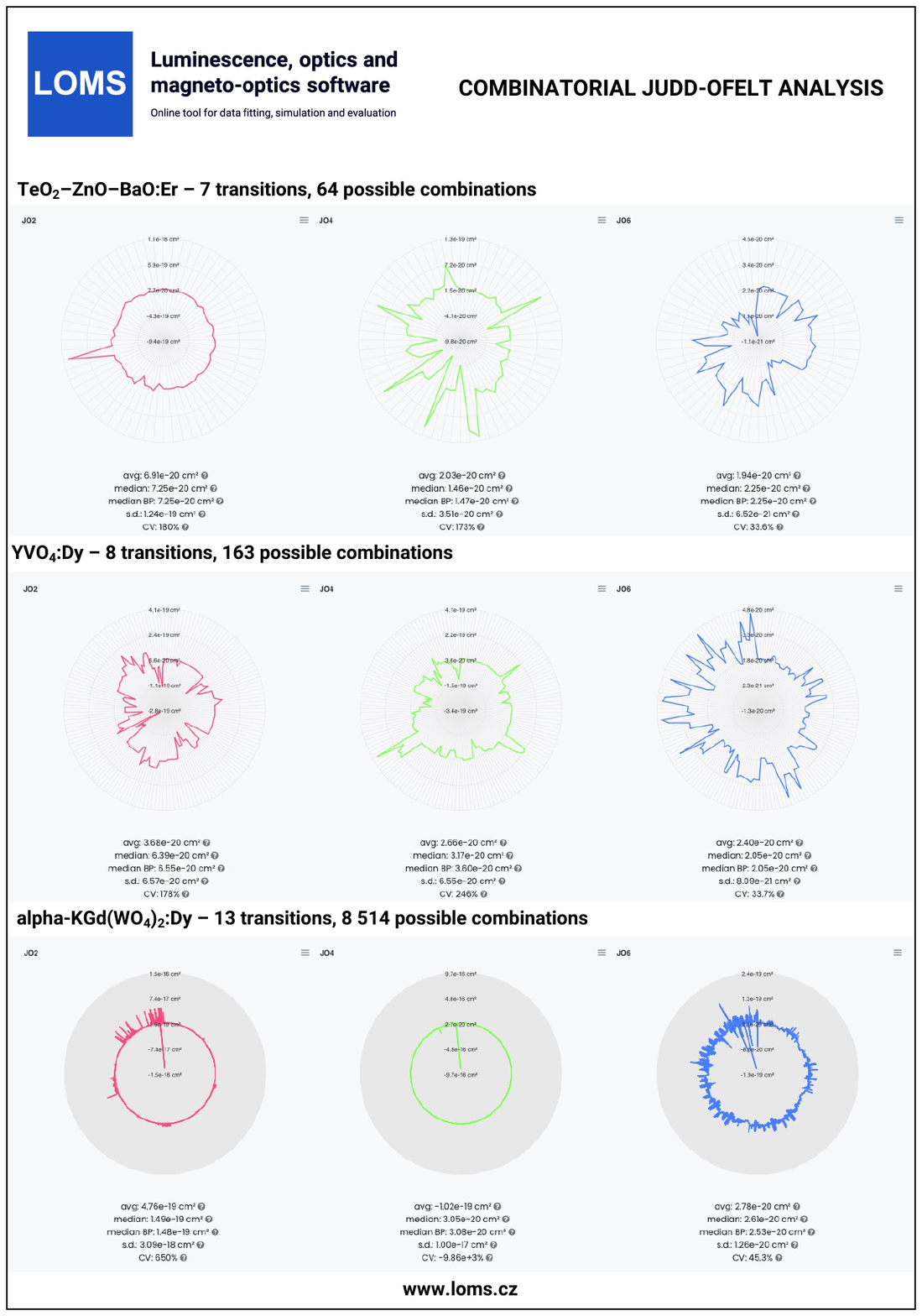}
\caption{\label{fig:jo_komb_1} Technical validation examples of combinatorial Judd-Ofelt analysis for materials doped with Er\textsuperscript{3+} and Dy\textsuperscript{3+} ions. Complete data outputs are listed in Figsahere repository\cite{redeposit}}
\end{figure}

\begin{figure}[h]
\centering
\includegraphics[width=13cm]{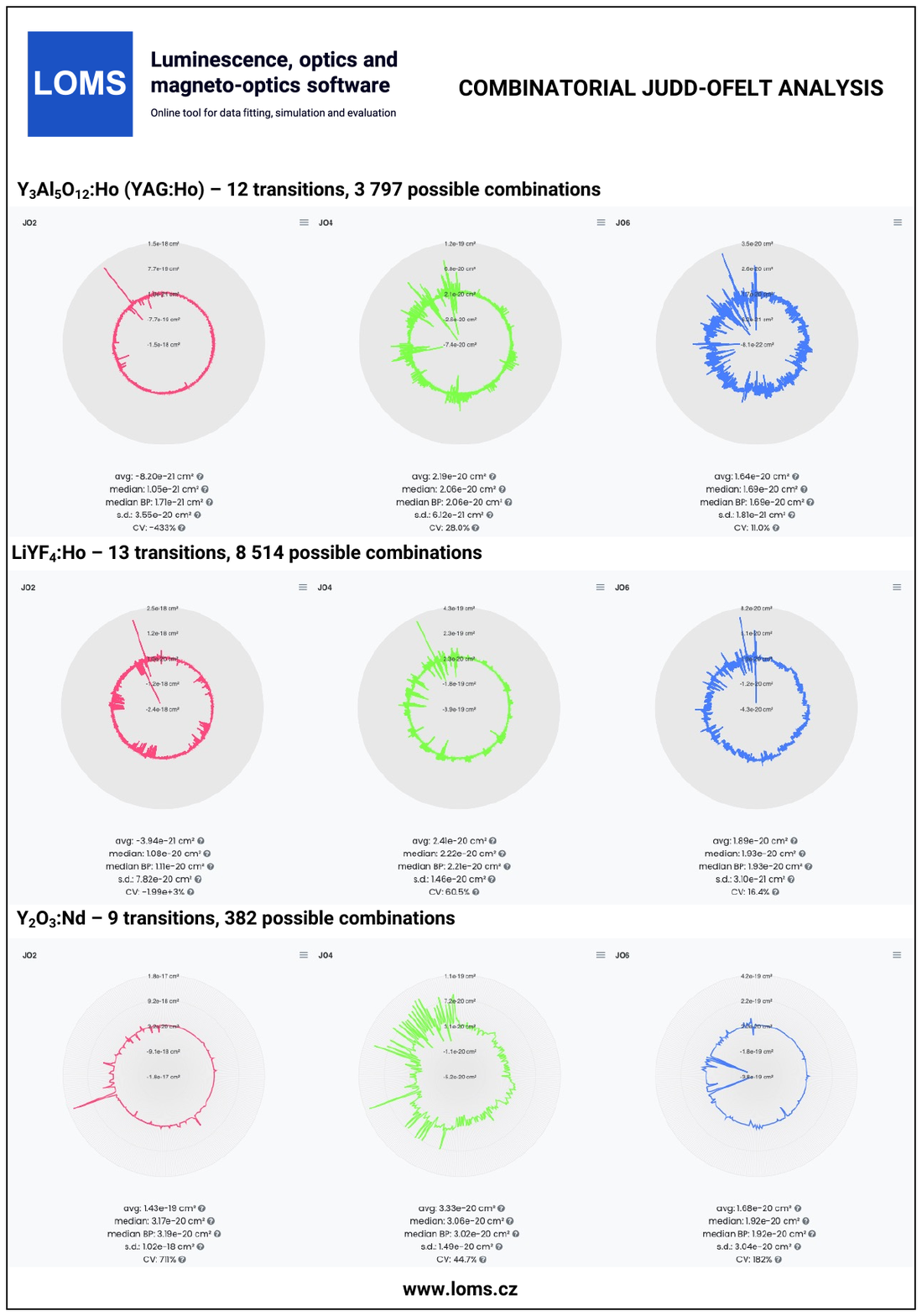}
\caption{\label{fig:jo_komb_2} Technical validation examples of combinatorial Judd-Ofelt analysis for materials doped with Ho\textsuperscript{3+} and Nd\textsuperscript{3+} ions. Complete data outputs are listed in Figsahere repository\cite{redeposit}}
\end{figure}

\begin{figure}[h]
\centering
\includegraphics[width=13cm]{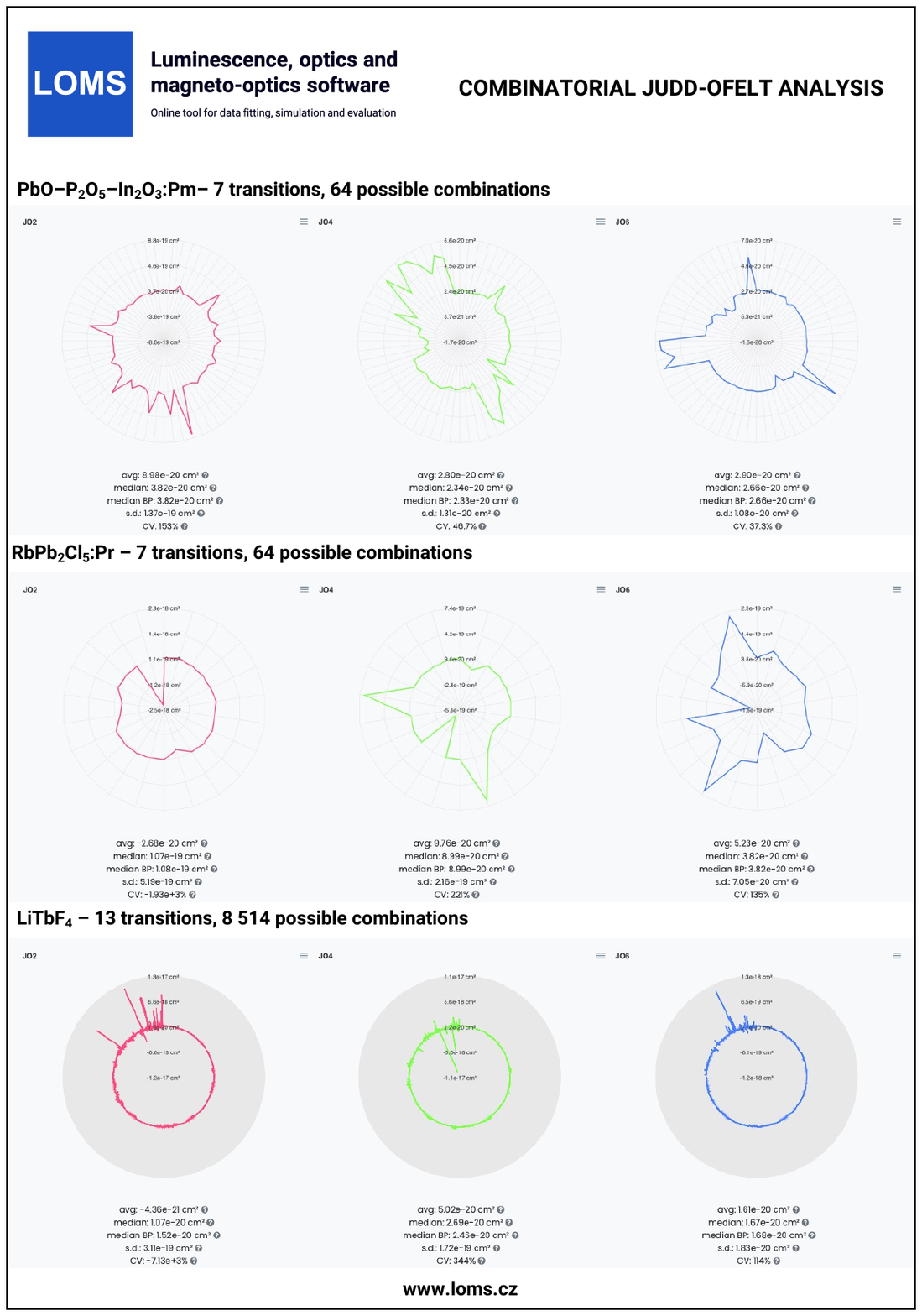}
\caption{\label{fig:jo_komb_3} Technical validation examples of combinatorial Judd-Ofelt analysis for materials doped with Pm\textsuperscript{3+}, Pr\textsuperscript{3+} and Tb\textsuperscript{3+} ions. Complete data outputs are listed in Figsahere repository\cite{redeposit}}
\end{figure}

\begin{figure}[h]
\centering
\includegraphics[width=13cm]{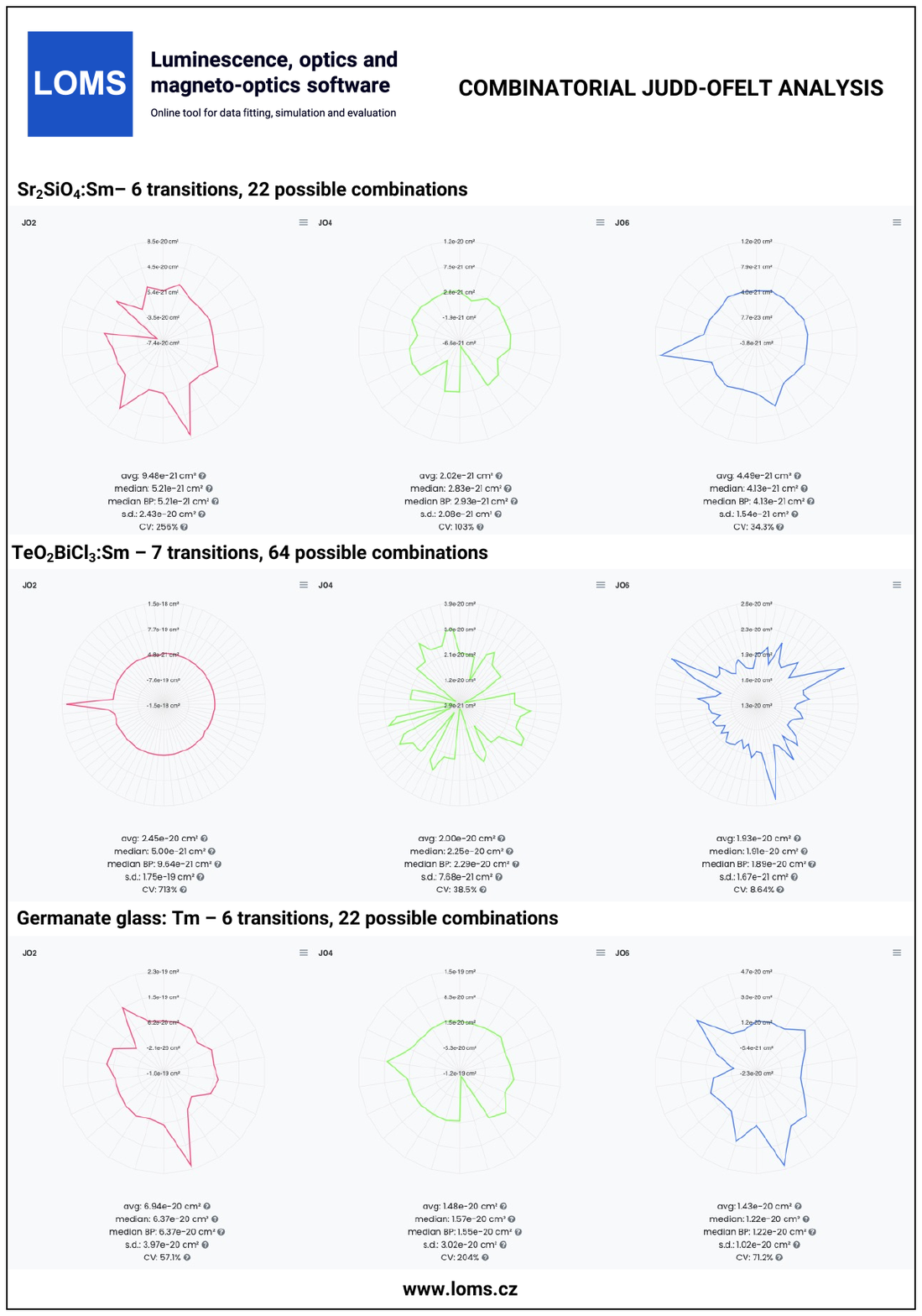}
\caption{\label{fig:jo_komb_4} Technical validation examples of combinatorial Judd-Ofelt analysis for materials doped with Sm\textsuperscript{3+} and Tm\textsuperscript{3+} ions. Complete data outputs are listed in Figshare repository\cite{redeposit}}
\end{figure}

The general procedure of JO analysis, radiative transition analysis and C-JO analysis is shown in a flow chart in Fig.\ref{fig:jo_flowchart} using four different input types: I. integrated cross section, II. experimental oscillator strength, III. experimental linestrength and IV. Judd-Ofelt parameters. Technical aspects and major steps in the process are described in the sections \textit{Judd-Ofelt theory: Experimental practice} and \textit{Evaluation protocol and graphical software interface}. The validity of the presented procedures is then presented in the following text on the examples of materials doped with Er\textsuperscript{3+}\cite{loms_0,loms_5}, Dy\textsuperscript{3+}\cite{loms_23,loms_86}, Ho\textsuperscript{3+}\cite{loms_1,loms_90}, Nd\textsuperscript{3+}\cite{loms_93}, Pm\textsuperscript{3+}\cite{loms_46}, Pr\textsuperscript{3+}\cite{loms_87}, Sm\textsuperscript{3+}\cite{loms_266,loms_72}, Tb\textsuperscript{3+}\cite{loms_88} and Tm\textsuperscript{3+}\cite{loms_68,loms_45} ions. Furthermore, C-JO analysis is provided for materials with more than four observed separate transitions, which allows the investigation of the most consistent and reliable outcomes using various combinations of absorption bands for JO analysis. Reference input files for all mentioned RE\textsuperscript{3+}-doped materials are included in Figshare data repository\cite{redeposit} together with a complete set of output files. The Combinatorial JO analysis results for selected RE\textsuperscript{3+} ions are presented within this text only in graphic form (Fig.\ref{fig:jo_komb_1}$-$\ref{fig:jo_komb_4}) due to the high number of possible combinations, where for 5, 6, 7, 8, 9, 10, 11, 12 and 13 experimentally observed input manifolds, it is possible to calculate 6, 22, 64, 163, 382, 848, 1816, 3797 and 8514 mutual manifolds combinations.  Complete step-by-step procedure is presented here for the first reference of TeO\textsubscript{2}$-$ZnO$-$BaO tellurite glass doped with Er\textsuperscript{3+} ions (TZB:Er)\cite{loms_0}. Other references are presented in shorter form concerning calculated JO parameters and results of the Combinatorial JO analysis (see Table\ref{tab:evaluation_param}). 

The Er-doped material (TeO\textsubscript{2}$-$ZnO$-$BaO glass) was chosen as the main example due to the presence of a reasonable number (seven observed manifolds) of absorption bands across the optical transmission spectral window when some of them may overlap with each other. The visible part of the absorption spectrum of TZB:Er glass is shown in Fig.\ref{fig:jo_porov_sigm}. Derived dependency of baseline corrected absorption cross section was used to obtain the integral in Eq.\ref{eq:jo_linestr}, which represents the integrated cross section (sum over the wavelength) for each observed manifold. These experimentally determined values were used as \textit{Input type I} in the LOMS.cz software accompanied by the positions of mean wavelength and refractive index value for each manifold to calculate the experimental linestrengths values, which were used for JO fitting. Figshare data repository also contains other possible input types formats for this material, where \textit{Input type II:} experimental oscillator strength, \textit{Input type III:} experimental linestrength and \textit{Input type IV:} JO parameters respectively. The last input type can be used together with known refractive index spectral dependency only for calcualtion of radiative properties. The placement of reduced matrix elements (U\textsuperscript{2}, U\textsuperscript{4}, U\textsuperscript{6}), integrated cross section, mean wavelength and both experimental and theoretical linestrength values within LOMS.cz GUI interface is shown in Fig.\ref{fig:jo_gui}. The JO parameters were found to be $\Omega_2 = 7.66\times10^{-20} $cm\textsuperscript{2}, $\Omega_4 = 1.51\times10^{-20} $cm\textsuperscript{2} and $\Omega_6 = 2.21\times10^{-20} $cm\textsuperscript{2} which is in agreement with values presented in Ref.\cite{loms_0} and values obtained by fitting procedure using the Walsh\cite{Walsh2006} evaluation software JOFwin(2011), where $\Omega_2 = 7.651\times10^{-20} $cm\textsuperscript{2}, $\Omega_4 = 1.508\times10^{-20} $cm\textsuperscript{2} and $\Omega_6 = 2.208\times10^{-20} $cm\textsuperscript{2}. The JO parameters were then used to calculate the transition probabilities according to Eq.\ref{eq:jo_transprob} between any excited state and any lower-lying energy level and to calculate the branching ratios and radiative lifetimes. The obtained results are shown in Fig.\ref{fig:loms_gui2} and Table \ref{tab:trans_er}. Note, that the data structure in Table \ref{tab:trans_er} is similar to the format of output file generated by LOMS online tool. The calculated values were again compared to those in Ref.\cite{loms_0} and calculated using JOFwin2011\cite{Walsh2006} software with a good agreement. It is thus possible to verify the validity and accuracy of JO analysis fitting procedure and calculation of radiative properties. To further verify the validity of LOMS.cz software calculations, a similar procedure was applied to other materials doped with RE\textsuperscript{3+} ions using different data Input types and various number of observed manifolds. Calculated results are listed in Table\ref{tab:evaluation_param} with corresponding references and denoted number of used manifolds in the parentheses. The presented results are in good agreement with the associated reference values and possible deviations are caused by the use of different values of reduced matrix elements, used constants and parameters or minor deviations in the calculation of linestrength values across the literature. 

The reference datasets with more than four manifolds were used for providing C-JO analysis and investigation of results consistency as the function of involved absorption bands in the calculation of JO parameters. The Table \ref{tab:evaluation_param} then contains the median values (Median) obtained from the set of all possible combinations and box-plot median values (Median BP)\cite{loms_0} obtained from the statistically reduced set of possible combinations which can be compared to the values of JO parameters calculated using the maximum possible number of observed manifolds (Full.set). Graphical results of C-JO analysis are shown in Fig.(\ref{fig:jo_komb_1}$-$\ref{fig:jo_komb_4}). The complete output files are included in Figshare repository\cite{redeposit} for detailed inspections.

\begin{table}[]
\caption{Calculated Judd-Ofelt radiative transition properties in TZB:Er glass using LOMS.cz online tool (in similar format as the software output file). The Transition eState represent the initial excited state ($J'$), Transition gState represent the final ground/lower lying state ($J$), $\lambda\textsubscript{em}$ is the emission wavelength calculated as the difference between involved energy levels which positions is represented by insterted values of Barycenters, S(ED) and S(MD) are electric and magnetic dipole line strengths and their respective contributions to the electric and magnetic transition probabilities A(ED) and A(MD), $\beta$ is the branching ratio and  last two columns represent the calculated values of radiative lifetime using the LOMS.cz online tool and those taken from Ref.\cite{loms_0}.}
\label{tab:trans_er}
\fontsize{8pt}{8pt}\selectfont
\begin{tabular}{llclllllcc}
\hline
Transition & Transition & $\lambda\textsubscript{em}$& S(ED)& S(MD)    & A(ED)    & A(MD)  & $\beta$       & $\tau$\textsuperscript{JO}\textsubscript{r} (LOMS)&$\tau$\textsuperscript{JO}\textsubscript{r} (Ref.\cite{loms_0})\\
 eState& gState& (nm)& \multicolumn{2}{c}{(cm\textsuperscript{2})}& \multicolumn{2}{c}{(s\textsuperscript{$-$1})}& & \multicolumn{2}{c}{(ms)}\\
\midrule
4I13/2            & 4I15/2            & 1547.7          & 3.49$\times10^{-20}$ & 7.04$\times10^{-21}$& 388      & 78.3   & 1.00       & 2.14           &2.15\\
 \hdashline
4I11/2            & 4I15/2            & 990.8           & 1.09$\times10^{-20}$ & 0.00     & 550      & 0.00   & 0.867      & 1.58           &1.58\\
4I11/2            & 4I13/2            & 2753.3          & 2.91$\times10^{-20}$ & 7.75$\times10^{-21}$& 66.5     & 17.7   & 0.133      & 11.9           &11.9\\
 \hdashline
4I9/2             & 4I15/2            & 812.8           & 2.84$\times10^{-21}$ & 0.00     & 316      & 0.00   & 0.626      & 1.98           &1.98\\
4I9/2             & 4I13/2            & 1711.7          & 1.60$\times10^{-20}$ & 0.00     & 184      & 0.00   & 0.364      & 5.30           &5.30\\
4I9/2             & 4I11/2            & 4524.9          & 4.06$\times10^{-21}$ & 3.91$\times10^{-21}$& 2.50     & 2.42   & 0.00973    & 204            &204\\
 \hdashline
4F9/2             & 4I15/2            & 661.0           & 1.83$\times10^{-20}$ & 0.00     & 3.90$\times10^{3}$  & 0.00   & 0.895      & 0.229          &0.230\\
4F9/2             & 4I13/2            & 1153.8          & 4.67$\times10^{-21}$ & 0.00     & 177      & 0.00   & 0.0407     & 2.18           &2.18\\
4F9/2             & 4I11/2            & 1986.1          & 3.39$\times10^{-20}$ & 1.86$\times10^{-21}$& 249      & 13.7   & 0.0602     & 3.55           &3.56\\
4F9/2             & 4I9/2             & 3539.8          & 1.05$\times10^{-20}$ & 4.39$\times10^{-21}$& 13.5     & 5.67   & 0.00440    & 52.1           &52.1\\
 \hdashline
4S3/2             & 4I15/2            & 547.9           & 4.89$\times10^{-21}$ & 0.00     & 4.77$\times10^{3}$  & 0.00   & 0.683      & 0.143          &0.144\\
4S3/2             & 4I13/2            & 848.2           & 7.65$\times10^{-21}$ & 0.00     & 1.87$\times10^{3}$  & 0.00   & 0.268      & 0.452          &0.453\\
4S3/2             & 4I11/2            & 1225.9          & 1.70$\times10^{-21}$ & 0.00     & 134      & 0.00   & 0.0192     & 2.92           &2.92\\
4S3/2             & 4I9/2             & 1681.5          & 6.81$\times10^{-21}$ & 0.00     & 206      & 0.00   & 0.0295     & 4.79           &4.80\\
4S3/2             & 4F9/2             & 3203.1          & 5.88$\times10^{-22}$ & 0.00     & 2.55     & 0.00   & 0.000366   & 392            &392\\
 \hdashline
2H11/2            & 4I15/2            & 526.3           & 6.29$\times10^{-20}$ & 0.00     & 2.33$\times10^{4}$  & 0.00   & 0.953      & 0.0408         &0.0409\\
2H11/2            & 4I13/2            & 797.4           & 3.85$\times10^{-21}$ & 1.41$\times10^{-21}$& 380      & 138    & 0.0212     & 0.873          &0.874\\
2H11/2            & 4I11/2            & 1122.6          & 5.64$\times10^{-21}$ & 5.08$\times10^{-22}$& 194      & 17.4   & 0.00864    & 1.59           &1.60\\
2H11/2            & 4I9/2             & 1493.0          & 2.32$\times10^{-20}$ & 9.50$\times10^{-23}$& 336      & 1.37   & 0.0138     & 2.41           &2.41\\
2H11/2            & 4F9/2             & 2582.0          & 2.82$\times10^{-20}$ & 1.10$\times10^{-22}$& 78.1     & 0.306  & 0.00320    & 12.7           &12.8\\
2H11/2            & 4S3/2             & 13315.6         & 3.23$\times10^{-21}$ & 0.00     & 0.0649   & 0.00   & 0.00000265 & 1.54$\times10^{4}$        &1.54$\times10^{4}$\\
 \hdashline
4F7/2             & 4I15/2            & 491.7           & 1.61$\times10^{-20}$ & 0.00     & 1.12$\times10^{4}$  & 0.00   & 0.838      & 0.0746         &0.0747\\
4F7/2             & 4I13/2            & 720.6           & 5.09$\times10^{-21}$ & 0.00     & 1.03$\times10^{3}$  & 0.00   & 0.0772     & 0.461          &0.462\\
4F7/2             & 4I11/2            & 976.0           & 7.62$\times10^{-21}$ & 0.00     & 604      & 0.00   & 0.0450     & 0.883          &0.884\\
4F7/2             & 4I9/2             & 1244.4          & 1.21$\times10^{-20}$ & 6.98$\times10^{-22}$& 458      & 26.3   & 0.0361     & 1.89           &1.89\\
4F7/2             & 4F9/2             & 1919.0          & 1.78$\times10^{-21}$ & 2.37$\times10^{-21}$& 18.1     & 24.1   & 0.00315    & 22.0           &22.0\\
4F7/2             & 4S3/2             & 4787.0          & 9.27$\times10^{-23}$ & 0.00     & 0.0603   & 0.00   & 0.00000450 & 311            &311\\
4F7/2             & 2H11/2            & 7473.8          & 1.85$\times10^{-20}$ & 0.00     & 3.16     & 0.00   & 0.000236   & 316            &317\\
\hline

\end{tabular}
\end{table}

\begin{table}[]
\caption{Comparison of the Judd–Ofelt parameters $\Omega$\textsubscript{i} (i = 2; 4; 6) for various materials with denoted number of involved manifolds for JO analysis in parenthesis. Calculated JO parameters were obtained using all experimentally measured manifolds (Full.set) or as a median value from a complete set (Median) or reduced set (by Box plot method - Median BP) of possible combinations calculated using Combinatorial Judd-Ofelt analysis.  }
\label{tab:evaluation_param}
\fontsize{8pt}{8pt}\selectfont

\begin{tabular}{ccllllc}
\hline
\multicolumn{1}{l}{RE\textsuperscript{3+}} & \multicolumn{1}{c}{Host matrix}                                                  & Involved & $\Omega_2$     & $\Omega_4$     & $\Omega_6$     & \multicolumn{1}{c}{Reference}    \\
 & & transitions& \multicolumn{3}{c}{($\times10^{-20} $cm\textsuperscript{2})}&\\ \hline
\multirow{5}{*}{Er\textsuperscript{3+}}    & \multirow{3}{*}{80TeO\textsubscript{2}$-$20ZnO$-$20BaO (glass)}                                 & Full.set (7)         & 7.66   & 1.51   & 2.21   & Hrabovsky (2024)\cite{loms_0}            \\
                         &                                                                             & Median               & 7.25   & 1.46   & 2.25   & and             \\ 
                         &                                                                             & Median BP            & 7.25   & 1.47   & 2.25   & This work              \\   \cline{3-7} 
                         & \multirow{2}{*}{Ge\textsubscript{25}-Ga\textsubscript{9.5}Sb\textsubscript{0.5}S\textsubscript{65} (glass)}                                 & Full.set (4)         & 4.31   & 2.46   & 1.96   & \multicolumn{1}{c}{Strizik (2014)\cite{loms_5}}      \\
                         &                                                                             & Full.set (4)         & 4.31   & 2.46   & 1.96   & \multicolumn{1}{c}{This work}    \\ \hline
\multirow{8}{*}{Dy\textsuperscript{3+}}    & \multirow{4}{*}{YVO\textsubscript{4} (single crystal)}                                      & Full.set (8)         & 6.59   & 3.71   & 1.74   & \multicolumn{1}{c}{Cavalli (2002)\cite{loms_23}}      \\ \cline{3-7} 
                         &                                                                             & Full.set (8)         & 6.56   & 3.60& 1.76   & \multirow{3}{*}{This work}       \\
                         &                                                                             & Median               & 6.39   & 3.17   & 2.05   &                                  \\
                         &                                                                             & Median BP            & 6.55   & 3.60& 2.05   &                                  \\ \cline{3-7} 
                         & \multirow{4}{*}{$\alpha -$KGd(WO\textsubscript{4})\textsubscript{2}}                                            & Full.set (13)        & 15.4& 3.05& 2.00& \multicolumn{1}{c}{Kaminskii (2002)\cite{loms_86}}    \\ \cline{3-7} 
                         &                                                                             & Full.set (13)        & 15.7   & 2.72   & 2.12   & \multirow{3}{*}{This work}       \\
                         &                                                                             & Median               & 14.9   & 3.05   & 2.61   &                                  \\
                         &                                                                             & Median BP            & 14.8   & 3.08   & 2.53   &                                  \\ \hline
\multirow{8}{*}{Ho\textsuperscript{3+}}      & \multirow{4}{*}{LiYF\textsubscript{4} (single crystal)}                                     & Full.set (13)        & 1.03   & 2.32   & 1.93   & \multicolumn{1}{c}{Walsh (1998)\cite{loms_90}}    \\ \cline{3-7} 
                         &                                                                             & Full.set (13)        & 1.03   & 2.31   & 1.94   & \multirow{3}{*}{This work}       \\
                         &                                                                             & Median               & 1.08   & 2.22   & 1.93   &                                  \\
                         &                                                                             & Median BP            & 1.11   & 2.21   & 1.93   &                                  \\ \cline{3-7} 
                         & \multirow{4}{*}{Y\textsubscript{3}Al\textsubscript{5}O\textsubscript{12} (single crystal)}                                  & Full.set (12)        & 0.10& 2.09& 1.72& \multicolumn{1}{c}{Walsh (2006)\cite{loms_1}} \\ \cline{3-7} 
                         &                                                                             & Full.set (12)        & 0.10& 2.08   & 1.73   & \multirow{3}{*}{This work}       \\
                         &                                                                             & Median               & 0.11& 2.06   & 1.69   &                                  \\
                         &                                                                             & Median BP            & 0.17& 2.06   & 1.69   &                                  \\ \hline
\multirow{4}{*}{Nd\textsuperscript{3+}}      & \multirow{4}{*}{Y\textsubscript{2}O\textsubscript{3}}                                                       & Full.set (9)         & 3.17& 3.08& 1.98& \multicolumn{1}{c}{Walsh (2002)\cite{loms_93}} \\ \cline{3-7} 
                         &                                                                             & Full.set (9)         & 3.17   & 3.09   & 1.99   & \multirow{3}{*}{This work}       \\
                         &                                                                             & Median               & 3.17   & 3.06   & 1.92   &                                  \\
                         &                                                                             & Median BP            & 3.19   & 3.02   & 1.92   &                                  \\ \hline
\multirow{4}{*}{Pm\textsuperscript{3+}}      & \multirow{4}{*}{65PbO-20P\textsubscript{2}O\textsubscript{5}-6In\textsubscript{2}O\textsubscript{3} (glass)}                           & Full.set (7)         & 3.80& 2.40& 2.6    & \multicolumn{1}{c}{Shinn (1988)\cite{loms_46}}        \\ \cline{3-7} 
                         &                                                                             & Full.set (7)         & 3.74   & 2.45   & 2.68   & \multirow{3}{*}{This work}       \\
                         &                                                                             & Median               & 3.82   & 2.34   & 2.66   &                                  \\
                         &                                                                             & Median BP            & 3.82   & 2.33   & 2.66   &                                  \\ \hline
\multirow{4}{*}{Pr\textsuperscript{3+}}      & \multicolumn{1}{c}{\multirow{4}{*}{RbPb\textsubscript{2}Cl\textsubscript{5}}}                               & Full.set (7)         & 10.6& 9.22   & 3.72   & \multicolumn{1}{c}{Merkle (2017)\cite{loms_87}}       \\ \cline{3-7} 
                         & \multicolumn{1}{l}{}                                                        & Full.set (7)         & 10.8   & 8.99   & 3.82   & \multirow{3}{*}{This work}       \\
                         & \multicolumn{1}{l}{}                                                        & Median               & 10.7   & 8.99   & 3.82   &                                  \\
                         & \multicolumn{1}{l}{}                                                        & Median BP            & 10.8   & 8.99   & 3.82   &                                  \\ \hline
\multirow{4}{*}{Tb\textsuperscript{3+}}      & \multirow{4}{*}{LiTbF\textsubscript{4}}                                                     & Full.set (13)        & 1.50& 2.23   & 2.06   & \multicolumn{1}{c}{Vasyliev (2013)\cite{loms_88}}     \\ \cline{3-7} 
                         &                                                                             & Full.set (13)        & 1.51   & 2.23   & 2.06   & \multirow{3}{*}{This work}       \\
                         &                                                                             & Median               & 1.07   & 2.69   & 1.67   &                                  \\
                         &                                                                             & Median BP            & 1.52   & 2.46   & 1.68   &                                  \\ \hline
\multirow{8}{*}{Sm\textsuperscript{3+}}      & \multicolumn{1}{c}{\multirow{4}{*}{Sr\textsubscript{2}SiO\textsubscript{4}}}                                & Full.set (6)         & 0.52   & 0.28& 0.40& \multicolumn{1}{c}{Manjunath (2018)\cite{loms_266}}    \\ \cline{3-7} 
                         & \multicolumn{1}{l}{}                                                        & Full.set (6)         & 0.57& 0.28& 0.40& \multirow{3}{*}{This work}       \\
                         & \multicolumn{1}{l}{}                                                        & Median               & 0.52& 0.28& 0.41&                                  \\
                         & \multicolumn{1}{l}{}                                                        & Median BP            & 0.52& 0.29& 0.41&                                  \\ \cline{3-7} 
                         & \multicolumn{1}{c}{\multirow{4}{*}{TeO\textsubscript{2}BiCl\textsubscript{3} (glass)}}                        & Full.set (7)         & 0.48   & 2.04   & 1.83   & \multicolumn{1}{c}{Boudchica (2023)\cite{loms_72}}    \\ \cline{3-7} 
                         & \multicolumn{1}{l}{}                                                        & Full.set (7)         & 0.48& 2.11   & 1.95   & \multirow{3}{*}{this work}       \\
                         & \multicolumn{1}{l}{}                                                        & Median               & 0.50& 2.25   & 1.91   &                                  \\
                         & \multicolumn{1}{l}{}                                                        & Median BP            & 0.96& 2.29   & 1.89   &                                  \\ \hline
\multirow{6}{*}{Tm\textsuperscript{3+}}      & \multicolumn{1}{l}{\multirow{4}{*}{GeO\textsubscript{2}–BaO/CaO–Na\textsubscript{2}O/Li\textsubscript{2}O (glass) }} & Full.set (6)         & 6.14   & 1.54   & 0.87   & \multicolumn{1}{c}{Walsh (2006)\cite{loms_68}}        \\ \cline{3-7} 
                         & \multicolumn{1}{l}{}                                                        & Full.set (6)         & 6.22   & 1.49   & 1.22   & \multirow{3}{*}{This work}       \\
                         & \multicolumn{1}{l}{}                                                        & Median               & 6.37   & 1.57   & 1.22   &                                  \\
                         & \multicolumn{1}{l}{}                                                        & Median BP            & 6.37   & 1.55   & 1.22   &                                  \\ \cline{3-7} 
                         & \multirow{2}{*}{Sr\textsubscript{5}(PO\textsubscript{4})\textsubscript{3}F (S-FAP crystal)}                                 & Full.set (4)         & 7.63& 10.5& 3.28& \multicolumn{1}{l}{Bonner (2006)\cite{loms_45}}       \\ \cline{3-7} 
                         &                                                                             & Full.set (4)         & 7.63   & 10.5   & 3.28   & \multicolumn{1}{c}{This work}    \\ \hline
\end{tabular}
\end{table}

\clearpage

\clearpage

\clearpage

\clearpage

\subsection*{Recommended output format and data reporting for Judd–Ofelt analysis}

To ensure reproducibility, comparability, and clarity in future research based on Judd–Ofelt analysis, we propose a standardized format for presenting calculated optical parameters and intermediate quantities. In many published studies, key details such as barycenter assignment, refractive indices, line strengths, or the units and conventions used for transition probabilities are inconsistently reported or omitted altogether. This hinders critical assessment and reuse of the data. In this section, we provide an example and recommendation for the minimal and extended information that should be included when reporting Judd–Ofelt parameters, radiative properties, and fitting quality. The proposed format aligns with the outputs generated by the LOMS.cz tool and supports the transparent dissemination of results necessary for scientific progress.

For proper presentation and reproducibility of results, all of the following parameters should be included:
\begin{enumerate}
  \item \textbf{Number and spectral parameters of involved transitions:} Clearly state the number of all observed and included transitions, along with the properly assigned values of the reduced matrix elements and their sources, unless explicitly listed elsewhere.
  \item \textbf{Individual transition characteristics:} Provide the spectral dependence of the refractive index and specify the main wavelength used, along with a description of how it was estimated. Provide the integrated absorption cross sections or integrated absorption coefficient values for the individual transitions in the supplementary section.
  \item \textbf{Oscillator strength / Line strength calculation:} Provide the exact procedure for both the experimental and theoretical determination of the relevant quantities - oscillator strength or line strength. Report the calculated values in commonly used units and appropriate orders of magnitude ($f\sim 10^{-6}; S\sim 10^{-20}$ cm$^{2}$). State, if the magnetic-dipole contribution was involved or not.
  \item \textbf{Judd-Ofelt parameters:} Provide all three Judd-Ofelt phenomenological parameters in commonly used units and appropriate orders of magnitude ($\Omega_2, \Omega_4, \Omega_6 \sim 10^{-20}$ cm$^{2}$) with corresponding uncertainties and quality fit parameter.
  \item \textbf{Radiative transition properties:} Provide the calculated values of transition probabilities, branching ratios, and Judd–Ofelt-calculated luminescence lifetimes, along with the corresponding values of the reduced matrix elements used or appropriate references to them. State, which mean wavelength values were used for calculation of aforementioned quantities.
\end{enumerate}

\subsection*{Code availability}
The complete set of blank template input files for each rare-earth ion, illustrative examples of input files together with attached results for JO and C-JO analysis and dataset of JO parameters listed in LOMS.cz database is available at Figshare\cite{redeposit} or the https://www.LOMS.cz/ webpage.

The LOMS.cz Software is licensed for personal, classroom, education and internal use only and not for the benefit of a third party (https://www.LOMS.cz/about/). The entire software codebase is publicly available on the LOMS.cz GitHub project (https://github.com/robinkrystufek/LOMS-JO). Presented repository of JO parameters is regularly updated to meet the ongoing scientific or industrial/engineering needs. Note that the data included in the JO parameters database and utilized in template/reference files are sourced from publicly available, peer-reviewed publications, such as scientific journals and handbooks/databooks. This curation ensures their reliability, and thus, their factual accuracy has not been further independently verified. Every data entry in the dataset or/and reference/template file clearly references its source,allowing users to explore the original data and its further context.
The Luminescence, Optics, and Magneto-Optics software (www.LOMS.cz) thus stands out as a vital resource by offering a user-friendly computational online tool for JO as well as C-JO analysis, and providing a comprehensive database of JO parameters in a standardized file format. With regular updates and open access, it proves indispensable for researchers, engineers, and students investigating the complex spectroscopic properties of rare-earth-doped materials.

\section*{Conclusion}
We have presented LOMS.cz, a comprehensive computational platform that addresses the long-standing challenge of standardizing Judd-Ofelt calculations in rare-earth spectroscopy. By integrating automated parameter computation, novel Combinatorial JO analysis, and a dynamically expanding database of validated parameters, LOMS.cz enables reliable extraction of spectroscopic properties while providing rigorous uncertainty quantification. The platform's capabilities have been extensively validated across diverse rare-earth systems including all spectroscopically active RE ions in various host matrices, demonstrating excellent agreement with established literature values.
Through its open-source nature and user-friendly interface, LOMS.cz establishes a foundation for accelerating the discovery and optimization of rare-earth-based photonic and optoelectronic materials. The platform's modular architecture supports continued expansion of its capabilities through community contributions, while its integrated database facilitates knowledge sharing and systematic comparison of results across different material systems. As the field continues to evolve, LOMS.cz provides a standardized framework that will enable researchers to efficiently evaluate materials properties and optimize rare-earth-doped systems for specific applications.

\section*{Acknowledgements}

J.H. gratefully acknowledges the support of the J.W. Fulbright Commission through the Fulbright-Masaryk Scholarship Program. This work was supported by the Czech Science Foundation, grants GA24-11361S, GA23-05507S and 24-10814S. Petr Vařák acknowledges the support of the junior grant awarded by the rector of UCT Prague.  Besides that, authors would like to thank Lukas Strizik for his valuable comments and discussion. Special thanks belong to Brian M. Walsh for his long-lasting and significant contribution to the field of RE ions spectroscopy and methodology of JO calculations which was an important factor in the development of this work.

\section*{Data availability statement}
Data is provided within the manuscript, supplementary information files or at Figshare redepository (10.6084/m9.figshare.c.7904465). Updated software documentation is available  at Luminescence, optics and magneto-optics software (LOMS) webpage https://www.LOMS.cz/ as well as the GUI of the software itself. The entire software codebase is publicly available on the LOMS.cz GitHub  project (https://github.com/robinkrystufek/LOMS-JO).

\section*{Author contributions statement}

All authors contributed equally to this work. J.H.: conceptualization, methodology, software, validation, formal analysis, investigation, visualization, supervision, funding aquisition and writing original draft, P.V. methodology, validation, formal analysis, visualization, investigation and writing - Review and editing; R.K. methodology, software, validation, formal analysis, visualization and writing - Review and editing. All authors reviewed the manuscript. 

\section*{Competing interest}
The authors declare no competing interests.

\section*{Additional information}
Correspondence and request for materials should be addressed to J. Hrabovsky.
Updated software documentation is available at Luminescence, optics and magneto-optics software (LOMS) webpage https://www.LOMS.cz/ as well as the GUI of the software itself.

\section*{Abbreviations}
\textbf{$A (J',J)$:} the probability of spontaneous emission for $J\rightarrow J'$ electric (\textbf{$A^{JO}_{ED}$}) or magnetic (\textbf{$A^{JO}_{MD}$}) dipole transitions;  \textbf{$\beta (J',J)$:} Branching ratio estimated from Judd-Ofelt analysis; $f^{abs}_{ED}:$ electric-dipole-induced oscillator strengths; $f^{abs}_{MD}:$ magnetic-dipole-induced oscillator strengths; $f^{abs}_{total}:$ total theoretical oscillator strength which contains both ED and MD contribution; \textit{g:} the electron g-factor; \textbf{$H$:} Hamiltonian; \textbf{$H_0$:} Hamiltonian representing the nucleus-electron interaction and the kinetic energies off all electrons; \textbf{$H\textsubscript{C}$:} Hamiltonian representing the Coulombic repulsion between electrons; \textbf{$H\textsubscript{F}$:} Hamiltonian of free ion; \textbf{$H\textsubscript{LF}$:} Perturbation Hamiltonian representing surrounding ligand/crystal field; \textbf{$H\textsubscript{SO}$:} Hamiltonian representing the spin-orbit interaction; \textit{h:} Planck's constant; $m_e$: electron mass; \textit{c}: speed of light in vacuum;  \textbf{$n_e$:} number of electrons in 4\textit{f} orbital; \textbf{$J$:} Total angular quantum momentum for ground state (J) and excited state (J’);  \textbf{$L$:} Total orbital quantum number; \textbf{$S$:} Total spin quantum number represented as the sum of electron spins; $S_{ED}:$ electric dipole line strength; $S_{MD}:$ magnetic dipole line strength; $^{2S+1}$L$_J$: Spectroscopic term symbol. Symbol \textit{S} denotes the total spin quantum number, $2S+1$ represents the spin multiplicity, \textit{L} is the total orbit quantum number and \textit{J} denotes the total angular quantum momentum; $Z$: atomic number; $\alpha_f$: fine structure constant; \textit{N:} rare-earth ion concentration; $\alpha(\lambda)$: wavelength dependent absorption coefficient; $\sigma_{abs}:$ absorption cross section; $\lambda$: wavelength; $\overline\lambda$: the mean wavelength of corresponding $J \rightarrow J'$; $\lambda_B$: transition wavelength - Barycenter;   $\chi_{ED}$: local field correction of the electric dipole; $\chi_{MD}$: local field correction of the electric dipole; $\tau\textsuperscript{JO}\textsubscript{r}$: calculated radiative lifetime from Judd-Ofelt analysis ; $\Omega_i$ ($i=2,4,6$): three Judd-Ofelt phenomenological parameters - $\Omega_2$, $\Omega_4$, $\Omega_6$; \textbf{C-JO:} Combinatorial Judd-ofelt; \textbf{ED:} electric-dipole; \textbf{MD:} magnetic-dipole; \textbf{IR:} Infrared part of electromagnetic spectrum; \textbf{JO:} Judd-Ofelt; \textbf{LOMS:} Luminescence, Optical and Magneto-optical Software, www.loms.cz; \textbf{MIR:} Mid-Infrared part of electromagnetic spectrum; \textbf{NIR:} Near-Infrared part of electromagnetic spectrum; \textbf{RE:} Rare-earth; \textbf{LED:} light-emitting diode; \textbf{UV:} Ultraviolet part of electromagnetic spectrum; \textbf{Vis:} Visible part of electromagnetic spectrum 

\bibliography{sample}

\begin{thebibliography}{10}
\urlstyle{rm}
\expandafter\ifx\csname url\endcsname\relax
  \def\url#1{\texttt{#1}}\fi
\expandafter\ifx\csname urlprefix\endcsname\relax\def\urlprefix{URL }\fi
\expandafter\ifx\csname doiprefix\endcsname\relax\def\doiprefix{DOI: }\fi
\providecommand{\bibinfo}[2]{#2}
\providecommand{\eprint}[2][]{\url{#2}}

\bibitem{WYBOURNE2004_re_intro}
\bibinfo{author}{Wybourne, B.}
\newblock \bibinfo{journal}{\bibinfo{title}{The fascination of the rare earths—then, now and in the future}}.
\newblock {\emph{\JournalTitle{J. Alloys Compd.}}} \textbf{\bibinfo{volume}{380}}, \bibinfo{pages}{96--100}, \url{10.1016/j.jallcom.2004.03.034} (\bibinfo{year}{2004}).

\bibitem{Walsh2006}
\bibinfo{author}{Walsh, B.}
\newblock \bibinfo{journal}{\bibinfo{title}{{ Judd-Ofelt theory: principles and practices}}}.
\newblock {\emph{\JournalTitle{Di Bartolo, B., Forte, O. (eds) Advances in Spectroscopy for Lasers and Sensing}}} \url{10.1007/1-4020-4789-4_21} (\bibinfo{year}{2006}).

\bibitem{chen_2017_RE}
\bibinfo{author}{Zhou, B.}, \bibinfo{author}{Li, Z.} \& \bibinfo{author}{Chen, C.}
\newblock \bibinfo{journal}{\bibinfo{title}{Global potential of rare earth resources and rare earth demand from clean technologies}}.
\newblock {\emph{\JournalTitle{Minerals}}} \textbf{\bibinfo{volume}{7}}, \url{10.3390/min7110203} (\bibinfo{year}{2017}).

\bibitem{liu_RE_spectroscopy}
\bibinfo{author}{Liu, G.} \& \bibinfo{author}{(eds), B.~J.}
\newblock \emph{\bibinfo{title}{Spectroscopic Properties of Rare Earths in Optical Materials}} (\bibinfo{publisher}{Springer}, \bibinfo{year}{2005}).

\bibitem{Baolu_Zhou_Rare_2016}
\bibinfo{author}{Zhou, B.}, \bibinfo{author}{Li, Z.}, \bibinfo{author}{Zhao, Y.}, \bibinfo{author}{Zhang, C.} \& \bibinfo{author}{Wei, Y.}
\newblock \bibinfo{journal}{\bibinfo{title}{Rare earth elements supply vs. clean energy technologies: new problems to be solve}}.
\newblock {\emph{\JournalTitle{Gospodarka Surowcami Mineralnymi - Mineral Resources Management}}} \textbf{\bibinfo{volume}{32}}, \url{10.1515/gospo-2016-0039} (\bibinfo{year}{2016}).

\bibitem{Gutfleisch_re_magnets}
\bibinfo{author}{Gutfleisch, O.} \emph{et~al.}
\newblock \bibinfo{journal}{\bibinfo{title}{Magnetic materials and devices for the 21st century: Stronger, lighter, and more energy efficient}}.
\newblock {\emph{\JournalTitle{Adv. Mater.}}} \textbf{\bibinfo{volume}{23}}, \bibinfo{pages}{821--842}, \url{10.1002/adma.201002180} (\bibinfo{year}{2011}).

\bibitem{sagawa_re_intro}
\bibinfo{author}{Sagawa, M.}, \bibinfo{author}{Fujimura, S.}, \bibinfo{author}{Togawa, N.}, \bibinfo{author}{Yamamoto, H.} \& \bibinfo{author}{Matsuura, Y.}
\newblock \bibinfo{journal}{\bibinfo{title}{{New material for permanent magnets on a base of Nd and Fe (invited)}}}.
\newblock {\emph{\JournalTitle{J. Appl. Phys.}}} \textbf{\bibinfo{volume}{55}}, \bibinfo{pages}{2083--2087}, \url{10.1063/1.333572} (\bibinfo{year}{1984}).

\bibitem{dong_re_intro}
\bibinfo{author}{Dong, H.} \emph{et~al.}
\newblock \bibinfo{journal}{\bibinfo{title}{Lanthanide nanoparticles: From design toward bioimaging and therapy}}.
\newblock {\emph{\JournalTitle{Chem. Rev.}}} \textbf{\bibinfo{volume}{115}}, \bibinfo{pages}{10725--10815}, \url{10.1021/acs.chemrev.5b00091} (\bibinfo{year}{2015}).

\bibitem{thou_re_intro_animalimag}
\bibinfo{author}{Zhou, J.}, \bibinfo{author}{Liu, Z.} \& \bibinfo{author}{Li, F.}
\newblock \bibinfo{journal}{\bibinfo{title}{Upconversion nanophosphors for small-animal imaging}}.
\newblock {\emph{\JournalTitle{Chem. Soc. Rev.}}} \textbf{\bibinfo{volume}{41}}, \bibinfo{pages}{1323--1349}, \url{10.1039/C1CS15187H} (\bibinfo{year}{2012}).

\bibitem{loms_market}
\bibinfo{author}{HCSS/TNO.}
\newblock \emph{\bibinfo{title}{Collaboration on Rare Earth Elements. An analysis of potentials for collaboration with Japan on Rare Earths}} (\bibinfo{publisher}{The Hague Centre for Strategic Studies and {TNO}: {The Hague, Netherlands}}, \bibinfo{year}{2012}).

\bibitem{loms_market2}
\bibinfo{author}{Tukker, A.}
\newblock \bibinfo{journal}{\bibinfo{title}{Rare earth elements supply restrictions: Market failures, not scarcity, hamper their current use in high-tech applications}}.
\newblock {\emph{\JournalTitle{Environmental Science {\&} Technology}}} \textbf{\bibinfo{volume}{48}}, \bibinfo{pages}{9973--9974}, \url{10.1021/es503548f} (\bibinfo{year}{2014}).

\bibitem{HEHLEN2013_juddofelt}
\bibinfo{author}{Hehlen, M.}, \bibinfo{author}{Brik, M.} \& \bibinfo{author}{Kramer, K.}
\newblock \bibinfo{journal}{\bibinfo{title}{50th anniversary of the {Judd–Ofelt} theory: {An} experimentalist's view of the formalism and its application}}.
\newblock {\emph{\JournalTitle{J. Lumin.}}} \textbf{\bibinfo{volume}{136}}, \bibinfo{pages}{221--239}, \url{10.1016/j.jlumin.2012.10.035} (\bibinfo{year}{2013}).

\bibitem{ciric2022_juddofelt}
\bibinfo{author}{Ciric, A.}, \bibinfo{author}{Marciniak, L.} \& \bibinfo{author}{Dramicanin, M.}
\newblock \bibinfo{journal}{\bibinfo{title}{Self-referenced method for the {Judd–Ofelt} parametrisation of the {Eu\textsuperscript{3+}} excitation spectrum.}}
\newblock {\emph{\JournalTitle{Sci Rep}}} \textbf{\bibinfo{volume}{12}}, \bibinfo{pages}{563}, \url{10.1038/s41598-021-04651-4} (\bibinfo{year}{2022}).

\bibitem{Judd1962}
\bibinfo{author}{Judd, B.}
\newblock \bibinfo{journal}{\bibinfo{title}{{Optical absorption intensities of rare-earth ions}}}.
\newblock {\emph{\JournalTitle{Physical Review}}} \textbf{\bibinfo{volume}{127}}, \bibinfo{pages}{750--761}, \url{10.1103/PhysRev.127.750} (\bibinfo{year}{1962}).

\bibitem{Ofelt962}
\bibinfo{author}{Ofelt, G.}
\newblock \bibinfo{journal}{\bibinfo{title}{{Intensities of Crystal Spectra of Rare‐Earth Ions}}}.
\newblock {\emph{\JournalTitle{The Journal of Chemical Physics}}} \textbf{\bibinfo{volume}{37}}, \bibinfo{pages}{511--520}, \url{10.1063/1.1701366} (\bibinfo{year}{1962}).

\bibitem{goldner_1996_jo}
\bibinfo{author}{Goldner, P.} \& \bibinfo{author}{Auzel, F.}
\newblock \bibinfo{journal}{\bibinfo{title}{{Application of standard and modified {Judd–Ofelt} theories to a praseodymium‐doped fluorozirconate glass}}}.
\newblock {\emph{\JournalTitle{J. Appl. Phys.}}} \textbf{\bibinfo{volume}{79}}, \bibinfo{pages}{7972--7977}, \url{10.1063/1.362347} (\bibinfo{year}{1996}).

\bibitem{smentek_2015_JO}
\bibinfo{author}{Smentek, L.}
\newblock \emph{\bibinfo{title}{Judd-Ofelt Theory — The Golden (and the Only One) Theoretical Tool of f-Electron Spectroscopy}}, chap.~\bibinfo{chapter}{10} (\bibinfo{publisher}{John Wiley {$\&$} Sons, Ltd}, \bibinfo{year}{2015}).

\bibitem{sytsma_1989_jo}
\bibinfo{author}{Sytsma, J.}, \bibinfo{author}{Imbusch, G.} \& \bibinfo{author}{Blasse, G.}
\newblock \bibinfo{journal}{\bibinfo{title}{{The spectroscopy of {Gd\textsuperscript{3+}} in yttriumoxychloride: {Judd–Ofelt} parameters from emission data}}}.
\newblock {\emph{\JournalTitle{J. Chem. Phys.}}} \textbf{\bibinfo{volume}{91}}, \bibinfo{pages}{1456--1461}, \url{10.1063/1.457106} (\bibinfo{year}{1989}).

\bibitem{loms_94}
\bibinfo{author}{G{\"o}rller-Walrand, C.} \& \bibinfo{author}{Binnemans, K.}
\newblock \bibinfo{title}{Chapter 167 spectral intensities of f-f transitions}.
\newblock In \emph{\bibinfo{booktitle}{Handbook on the Physics and Chemistry of Rare Earths}}, \bibinfo{pages}{101--264}, \url{10.1016/S0168-1273(98)25006-9} (\bibinfo{publisher}{Elsevier}, \bibinfo{year}{1998}).

\bibitem{CIRIC2019116749}
\bibinfo{author}{Ciric, A.}, \bibinfo{author}{Stojadinovic, S.} \& \bibinfo{author}{Dramicanin, M.}
\newblock \bibinfo{journal}{\bibinfo{title}{An extension of the judd-ofelt theory to the field of lanthanide thermometry}}.
\newblock {\emph{\JournalTitle{J. Lumin.}}} \textbf{\bibinfo{volume}{216}}, \bibinfo{pages}{116749}, \url{10.1016/j.jlumin.2019.116749} (\bibinfo{year}{2019}).

\bibitem{CIRIC2019395_jo_intro}
\bibinfo{author}{Ciric, A.}, \bibinfo{author}{Stojadinovic, S.} \& \bibinfo{author}{Dramicanin, M.}
\newblock \bibinfo{journal}{\bibinfo{title}{Approximate prediction of the cie coordinates of lanthanide-doped materials from the judd-ofelt intensity parameters}}.
\newblock {\emph{\JournalTitle{J. Lumin.}}} \textbf{\bibinfo{volume}{213}}, \bibinfo{pages}{395--400}, \url{10.1016/j.jlumin.2019.05.052} (\bibinfo{year}{2019}).

\bibitem{Ciric2019}
\bibinfo{author}{Ciric, A.}, \bibinfo{author}{Stojadinovic, S.}, \bibinfo{author}{Sekulic, M.} \& \bibinfo{author}{Dramicanin, M.}
\newblock \bibinfo{journal}{\bibinfo{title}{{JOES: An application software for Judd-Ofelt analysis from {Eu\textsuperscript{3+}} emission spectra}}}.
\newblock {\emph{\JournalTitle{Journal of Luminescence}}} \textbf{\bibinfo{volume}{205}}, \url{10.1016/j.jlumin.2018.09.048} (\bibinfo{year}{2019}).

\bibitem{Dutra2014}
\bibinfo{author}{Dutra, J.}, \bibinfo{author}{Bispo, T.} \& \bibinfo{author}{Freire, R.}
\newblock \bibinfo{journal}{\bibinfo{title}{{LUMPAC lanthanide luminescence software: Efficient and user friendly}}}.
\newblock {\emph{\JournalTitle{Journal of Computational Chemistry}}} \textbf{\bibinfo{volume}{35}}, \url{10.1002/jcc.23542} (\bibinfo{year}{2014}).

\bibitem{Moura2021}
\bibinfo{author}{Moura~Jr., R.} \emph{et~al.}
\newblock \bibinfo{journal}{\bibinfo{title}{{(INVITED) JOYSpectra: A web platform for luminescence of lanthanides}}}.
\newblock {\emph{\JournalTitle{Journal of Computational Chemistry}}} \textbf{\bibinfo{volume}{11}}, \url{10.1016/j.omx.2021.100080} (\bibinfo{year}{2021}).

\bibitem{Hrabovsky_2021_yag_luag}
\bibinfo{author}{Hrabovsk\'{y}, J.}, \bibinfo{author}{Ku\v{c}era, M.}, \bibinfo{author}{Palou\v{s}ov\'{a}, L.}, \bibinfo{author}{Bi, L.} \& \bibinfo{author}{Veis, M.}
\newblock \bibinfo{journal}{\bibinfo{title}{Optical characterization of {Y\textsubscript{3}Al\textsubscript{5}O\textsubscript{12}} and {Lu\textsubscript{3}Al\textsubscript{5}O\textsubscript{12}} single crystals}}.
\newblock {\emph{\JournalTitle{Opt. Mater. Express}}} \textbf{\bibinfo{volume}{11}}, \bibinfo{pages}{1218--1223}, \url{10.1364/OME.417670} (\bibinfo{year}{2021}).

\bibitem{Hrabovsky_2024_RE_mapping}
\bibinfo{author}{Hrabovsk\'{y}, J.} \emph{et~al.}
\newblock \bibinfo{journal}{\bibinfo{title}{{Rapid and precise large area mapping of rare-earth doping homogeneity in luminescent materials}}}.
\newblock {\emph{\JournalTitle{Commun Mater}}} \textbf{\bibinfo{volume}{5}}, \bibinfo{pages}{251}, \url{10.1038/s43246-024-00679-x} (\bibinfo{year}{2024}).

\bibitem{Dieke_1963_jo}
\bibinfo{author}{Dieke, G.} \& \bibinfo{author}{Crosswhite, H.}
\newblock \bibinfo{journal}{\bibinfo{title}{The spectra of the doubly and triply ionized rare earths}}.
\newblock {\emph{\JournalTitle{Appl. Opt.}}} \textbf{\bibinfo{volume}{2}}, \bibinfo{pages}{675--686}, \url{10.1364/AO.2.000675} (\bibinfo{year}{1963}).

\bibitem{PEIJZEL2005_jo}
\bibinfo{author}{Peijzel, P.}, \bibinfo{author}{Meijerink, A.}, \bibinfo{author}{Wegh, R.}, \bibinfo{author}{Reid, M.} \& \bibinfo{author}{Burdick, G.}
\newblock \bibinfo{journal}{\bibinfo{title}{A complete 4fn energy level diagram for all trivalent lanthanide ions}}.
\newblock {\emph{\JournalTitle{J. Sol. State Chem.}}} \textbf{\bibinfo{volume}{178}}, \bibinfo{pages}{448--453}, \url{10.1016/j.jssc.2004.07.046} (\bibinfo{year}{2005}).

\bibitem{vleck_1937}
\bibinfo{author}{Vleck, J.}
\newblock \bibinfo{journal}{\bibinfo{title}{The puzzle of rare-earth spectra in solids.}}
\newblock {\emph{\JournalTitle{J. Phys. Chem.}}} \textbf{\bibinfo{volume}{41}}, \bibinfo{pages}{67--80}, \url{10.1021/j150379a006} (\bibinfo{year}{1937}).

\bibitem{BROER_1945_JO}
\bibinfo{author}{Broer, L.}, \bibinfo{author}{Gorter, C.} \& \bibinfo{author}{Hoogschagen, J.}
\newblock \bibinfo{journal}{\bibinfo{title}{On the intensities and the multipole character in the spectra of the rare earth ions}}.
\newblock {\emph{\JournalTitle{Physica}}} \textbf{\bibinfo{volume}{11}}, \bibinfo{pages}{231--250}, \url{10.1016/S0031-8914(45)80009-5} (\bibinfo{year}{1945}).

\bibitem{edgar_fmd_2006}
\bibinfo{author}{Edgar, A.}
\newblock \emph{\bibinfo{title}{Optical properties of glasses, in. J. Singh (Ed.), Optical Properties of Condensed Matter and Applications}} (\bibinfo{publisher}{John Wiley {$\&$} Sons}, \bibinfo{year}{2006}).

\bibitem{wybourne_1965}
\bibinfo{author}{Wybourne, B.}
\newblock \emph{\bibinfo{title}{Spectroscopic properties of Rare Earths}} (\bibinfo{publisher}{Wiley (New York)}, \bibinfo{year}{1965}).

\bibitem{loms_0}
\bibinfo{author}{Hrabovsky, J.} \emph{et~al.}
\newblock \bibinfo{journal}{\bibinfo{title}{Classical and combinatorial {Judd–Ofelt} analysis of spectroscopic properties in {Er-doped materials: TeO\textsubscript{2}-ZnO-BaO:Er\textsuperscript{3+}}: glasses}}.
\newblock {\emph{\JournalTitle{J. Phys. Photonics}}} \textbf{\bibinfo{volume}{7}}, \bibinfo{pages}{025006}, \url{10.1088/2515-7647/adb115} (\bibinfo{year}{2025}).

\bibitem{Carnall1968_error1}
\bibinfo{author}{Carnall, W.}, \bibinfo{author}{Fields, P.} \& \bibinfo{author}{Rajnak, K.}
\newblock \bibinfo{journal}{\bibinfo{title}{Electronic energy levels in the trivalent lanthanide aquo ions. {I. Pr\textsuperscript{3+}, Nd\textsuperscript{3+}, Pm\textsuperscript{3+}, Sm\textsuperscript{3+}, Dy\textsuperscript{3+}, Ho\textsuperscript{3+}, Er\textsuperscript{3+}, and Tm\textsuperscript{3+}}}}.
\newblock {\emph{\JournalTitle{J. Chem. Phys.}}} \textbf{\bibinfo{volume}{49}}, \bibinfo{pages}{4424--4442}, \url{10.1063/1.1669893} (\bibinfo{year}{1968}).

\bibitem{Zhang2020_error2}
\bibinfo{author}{Zhang, Y.} \emph{et~al.}
\newblock \bibinfo{journal}{\bibinfo{title}{Error evaluation of {Judd-Ofelt} spectroscopic analysis}}.
\newblock {\emph{\JournalTitle{Spectrochim. Acta A Mol. Biomol. Spectrosc.}}} \textbf{\bibinfo{volume}{239}}, \bibinfo{pages}{118536}, \url{10.1016/j.saa.2020.118536} (\bibinfo{year}{2020}).

\bibitem{Afef_2018_hyper1}
\bibinfo{author}{Afef, B.} \emph{et~al.}
\newblock \bibinfo{journal}{\bibinfo{title}{Green and near-infrared emission of {Er\textsuperscript{3+} doped PZS and PZC glasses}}}.
\newblock {\emph{\JournalTitle{J. Lumin.}}} \textbf{\bibinfo{volume}{194}}, \bibinfo{pages}{706–712}, \url{10.1016/j.jlumin.2017.09.040} (\bibinfo{year}{2018}).

\bibitem{Karthikeyan_2015_hyper2}
\bibinfo{author}{Karthikeyan, P.}, \bibinfo{author}{Suthanthirakumar, P.}, \bibinfo{author}{Vijayakumar, R.} \& \bibinfo{author}{Marimuthu, K.}
\newblock \bibinfo{journal}{\bibinfo{title}{Structural and luminescence behaviour of {Er\textsuperscript{3+}} doped telluro-fluoroborate glasses}}.
\newblock {\emph{\JournalTitle{J. Mol. Struct.}}} \textbf{\bibinfo{volume}{1083}}, \bibinfo{pages}{268--277}, \url{10.1016/j.molstruc.2014.12.003} (\bibinfo{year}{2015}).

\bibitem{Mariyappan_2019_hyper3}
\bibinfo{author}{Mariyappan, M.}, \bibinfo{author}{Arunkumar, S.} \& \bibinfo{author}{Marimuthu, K.}
\newblock \bibinfo{journal}{\bibinfo{title}{{Judd-Ofelt analysis and {NIR} luminescence investigations on {Er\textsuperscript{3+}} ions doped {B\textsubscript{2}O\textsubscript{3}--Bi\textsubscript{2}O\textsubscript{3}--Li\textsubscript{2}O--K\textsubscript{2}O} glasses for photonic applications}}}.
\newblock {\emph{\JournalTitle{Physica B Condens. Matter}}} \textbf{\bibinfo{volume}{572}}, \bibinfo{pages}{27--35}, \url{10.1016/j.physb.2019.07.036} (\bibinfo{year}{2019}).

\bibitem{Aoki_2017_trans0}
\bibinfo{author}{Aoki, T.}, \bibinfo{author}{Strizik, L.}, \bibinfo{author}{Hrabovsky, J.} \& \bibinfo{author}{Wagner, T.}
\newblock \bibinfo{journal}{\bibinfo{title}{Quadrature frequency resolved spectroscopy of upconversion photoluminescence in {GeGaS:Er\textsuperscript{3+}}; {II}. elucidating excitation mechanisms of red emission besides green emission}}.
\newblock {\emph{\JournalTitle{J. Mater. Sci.: Mater. Electron.}}} \textbf{\bibinfo{volume}{28}}, \bibinfo{pages}{7077--7082}, \url{10.1007/s10854-017-6363-2} (\bibinfo{year}{2017}).

\bibitem{Bruno_Bureau_2009_trans1}
\bibinfo{author}{Bureau, B.} \emph{et~al.}
\newblock \bibinfo{journal}{\bibinfo{title}{Chalcogenide glass fibers for infrared sensing and space optics}}.
\newblock {\emph{\JournalTitle{Fiber Integr. Opt.}}} \textbf{\bibinfo{volume}{28}}, \bibinfo{pages}{65--80}, \url{10.1080/01468030802272542} (\bibinfo{year}{2009}).

\bibitem{Seddon1995_trans2}
\bibinfo{author}{Seddon, A.}
\newblock \bibinfo{journal}{\bibinfo{title}{Chalcogenide glasses: a review of their preparation, properties and applications}}.
\newblock {\emph{\JournalTitle{J. Non Cryst. Solids}}} \textbf{\bibinfo{volume}{184}}, \bibinfo{pages}{44--50}, \url{10.1016/0022-3093(94)00686-5} (\bibinfo{year}{1995}).

\bibitem{refr_index_info}
\bibinfo{author}{Polyanskiy, M.}
\newblock \bibinfo{journal}{\bibinfo{title}{{Refractiveindex.info database of optical constants}}}.
\newblock {\emph{\JournalTitle{Sci data}}} \textbf{\bibinfo{volume}{11}}, \url{10.1038/s41597-023-02898-2} (\bibinfo{year}{2024}).

\bibitem{redeposit}
\bibinfo{author}{Hrabovsky, J.}, \bibinfo{author}{Varak, P.} \& \bibinfo{author}{Krystufek, R.}
\newblock \bibinfo{journal}{\bibinfo{title}{{LOMS.cz: interactive online software for Classical and Combinatorial Judd-Ofelt analysis with integrated database of Judd-Ofelt parameters}}}.
\newblock {\emph{\JournalTitle{Figshare}}} \url{10.6084/m9.figshare.c.7904465} (\bibinfo{year}{2024}).

\bibitem{loms_87}
\bibinfo{author}{Merkle, L.~D.} \& \bibinfo{author}{Dubinskii, M.}
\newblock \bibinfo{journal}{\bibinfo{title}{{Pr:RbPb\textsubscript{2}Cl\textsubscript{5}}: temperature dependent spectra, dynamics and three-for-one excitation}}.
\newblock {\emph{\JournalTitle{Opt. Express}}} \textbf{\bibinfo{volume}{25}}, \bibinfo{pages}{19780}, \url{10.1364/OE.25.019780} (\bibinfo{year}{2017}).

\bibitem{loms_93}
\bibinfo{author}{Walsh, B.~M.} \emph{et~al.}
\newblock \bibinfo{journal}{\bibinfo{title}{Spectroscopic characterization of {Nd:Y\textsubscript{2}O\textsubscript{3}}: application toward a differential absorption lidar system for remote sensing of ozone}}.
\newblock {\emph{\JournalTitle{J. Opt. Soc. Am. B}}} \textbf{\bibinfo{volume}{19}}, \bibinfo{pages}{2893}, \url{10.1364/JOSAB.19.002893} (\bibinfo{year}{2002}).

\bibitem{loms_46}
\bibinfo{author}{Shinn, M.~D.}, \bibinfo{author}{Krupke, W.~F.}, \bibinfo{author}{Solarz, R.~W.} \& \bibinfo{author}{Kirchoff, T.~A.}
\newblock \bibinfo{journal}{\bibinfo{title}{Spectroscopic and laser properties of {Pm\textsuperscript{3+}}}}.
\newblock {\emph{\JournalTitle{IEEE J. Quantum Electron.}}} \textbf{\bibinfo{volume}{24}}, \bibinfo{pages}{1100--1108}, \url{10.1109/3.232} (\bibinfo{year}{1988}).

\bibitem{loms_266}
\bibinfo{author}{Manjunath, C.} \emph{et~al.}
\newblock \bibinfo{journal}{\bibinfo{title}{Optical absorption intensity analysis using judd-ofelt theory and photoluminescence investigation of orange-red {Sr\textsubscript{2}SiO\textsubscript{4}}: {Sm\textsuperscript{3+}} nanopigments}}.
\newblock {\emph{\JournalTitle{Dyes and Pigments}}} \textbf{\bibinfo{volume}{148}}, \bibinfo{pages}{118--129}, \url{10.1016/j.dyepig.2017.08.036} (\bibinfo{year}{2018}).

\bibitem{loms_72}
\bibinfo{author}{Boudchicha, N.} \emph{et~al.}
\newblock \bibinfo{journal}{\bibinfo{title}{{Judd-Ofelt} analysis and spectroscopy study of tellurite glasses doped with rare-earth ({Nd\textsuperscript{3+}}, {Sm\textsuperscript{3+}}, {Dy\textsuperscript{3+}}, and {Er\textsuperscript{3+}})}}.
\newblock {\emph{\JournalTitle{Materials (Basel)}}} \textbf{\bibinfo{volume}{16}}, \bibinfo{pages}{6832}, \url{10.3390/ma16216832} (\bibinfo{year}{2023}).

\bibitem{loms_88}
\bibinfo{author}{Vasyliev, V.}, \bibinfo{author}{Villora, E.~G.}, \bibinfo{author}{Sugahara, Y.} \& \bibinfo{author}{Shimamura, K.}
\newblock \bibinfo{journal}{\bibinfo{title}{{Judd-Ofelt} analysis and emission quantum efficiency of {Tb}-fluoride single crystals: {LiTbF\textsubscript{4}} and {Tb\textsubscript{0.81}Ca\textsubscript{0.19}F\textsubscript{42.81}}}}.
\newblock {\emph{\JournalTitle{J. Appl. Phys.}}} \textbf{\bibinfo{volume}{113}}, \bibinfo{pages}{203508}, \url{10.1063/1.4807649} (\bibinfo{year}{2013}).

\bibitem{loms_23}
\bibinfo{author}{Cavalli, E.}, \bibinfo{author}{Bettinelli, M.}, \bibinfo{author}{Belletti, A.} \& \bibinfo{author}{Speghini, A.}
\newblock \bibinfo{journal}{\bibinfo{title}{Optical spectra of yttrium phosphate and yttrium vanadate single crystals activated with {Dy\textsuperscript{3+}}}}.
\newblock {\emph{\JournalTitle{J. Alloys Compd.}}} \textbf{\bibinfo{volume}{341}}, \bibinfo{pages}{107--110}, \url{10.1016/S0925-8388(02)00079-8} (\bibinfo{year}{2002}).

\bibitem{loms_86}
\bibinfo{author}{Kaminskii, A.} \emph{et~al.}
\newblock \bibinfo{journal}{\bibinfo{title}{Optical spectroscopy and visible stimulated emission of {Dy\textsuperscript{3+}} ions in {monoclinic $\alpha-$KY(WO\textsubscript{4})\textsubscript{2} and $\alpha-$KGd(WO\textsubscript{4})\textsubscript{2} crystals}}}.
\newblock {\emph{\JournalTitle{Phys. Rev. B Condens. Matter}}} \textbf{\bibinfo{volume}{65}}, \url{10.1103/PhysRevB.65.125108} (\bibinfo{year}{2002}).

\bibitem{loms_1}
\bibinfo{author}{Walsh, B.}, \bibinfo{author}{Grew, G.} \& \bibinfo{author}{Barnes, N.}
\newblock \bibinfo{journal}{\bibinfo{title}{Energy levels and intensity parameters of ions in {Y\textsubscript{3}Al\textsubscript{5}O\textsubscript{12}} and {Lu\textsubscript{3}Al\textsubscript{5}O\textsubscript{12}}}}.
\newblock {\emph{\JournalTitle{J. Phys. Chem. Solids}}} \textbf{\bibinfo{volume}{67}}, \bibinfo{pages}{1567--1582}, \url{10.1016/j.jpcs.2006.01.123} (\bibinfo{year}{2006}).

\bibitem{loms_90}
\bibinfo{author}{Walsh, B.~M.}, \bibinfo{author}{Barnes, N.~P.} \& \bibinfo{author}{Di~Bartolo, B.}
\newblock \bibinfo{journal}{\bibinfo{title}{Branching ratios, cross sections, and radiative lifetimes of rare earth ions in solids: Application to {Tm\textsuperscript{3+}} and {Ho\textsuperscript{3+}} ions in {LiYF\textsubscript{4}}}}.
\newblock {\emph{\JournalTitle{J. Appl. Phys.}}} \textbf{\bibinfo{volume}{83}}, \bibinfo{pages}{2772--2787}, \url{10.1063/1.367037} (\bibinfo{year}{1998}).

\bibitem{loms_5}
\bibinfo{author}{Strizik, L.} \emph{et~al.}
\newblock \bibinfo{journal}{\bibinfo{title}{Green, red and near-infrared photon up-conversion in {Ga--Ge--Sb--S:{Er\textsuperscript{3+}}} amorphous chalcogenides}}.
\newblock {\emph{\JournalTitle{J. Lumin.}}} \textbf{\bibinfo{volume}{147}}, \bibinfo{pages}{209--215}, \url{10.1016/j.jlumin.2013.11.021} (\bibinfo{year}{2014}).

\bibitem{loms_68}
\bibinfo{author}{Walsh, B.~M.}, \bibinfo{author}{Barnes, N.~P.}, \bibinfo{author}{Reichle, D.~J.} \& \bibinfo{author}{Jiang, S.}
\newblock \bibinfo{journal}{\bibinfo{title}{Optical properties of {Tm\textsuperscript{3+}} ions in alkali germanate glass}}.
\newblock {\emph{\JournalTitle{J. Non Cryst. Solids}}} \textbf{\bibinfo{volume}{352}}, \bibinfo{pages}{5344--5352}, \url{10.1016/j.jnoncrysol.2006.08.029} (\bibinfo{year}{2006}).

\bibitem{loms_45}
\bibinfo{author}{Bonner, C.} \emph{et~al.}
\newblock \bibinfo{journal}{\bibinfo{title}{A spectroscopic and {Judd--Ofelt} analysis of the relaxation dynamics of {Tm\textsuperscript{3+}} in the fluorapatites, {FAP}, {S-FAP}, and {B-FAP}}}.
\newblock {\emph{\JournalTitle{Opt. Mater. (Amst.)}}} \textbf{\bibinfo{volume}{20}}, \bibinfo{pages}{1--12}, \url{10.1016/S0925-3467(02)00005-8} (\bibinfo{year}{2002}).

\end{thebibliography}

\end{document}